\documentclass[preprint]{aastex}
\shorttitle{High-Ionization Emission in Metal-Deficient BCDs}
\shortauthors{T. X. Thuan \& Y. I. Izotov}

\begin{document}

\title{High-Ionization Emission in Metal-Deficient Blue Compact Dwarf 
Galaxies\footnote{The observations reported here were obtained at the MMT 
Observatory, a joint facility of the Smithsonian Institution and the University of Arizona.}}

\author{Trinh X. Thuan}
\affil{Astronomy Department, University of Virginia,
    Charlottesville, VA 22903}
\email{txt@virginia.edu}

\and

\author{Yuri I. Izotov}
\affil{Main Astronomical Observatory, National Academy of Sciences of Ukraine, 03680, Kyiv, Ukraine}
\email{izotov@mao.kiev.ua}

\begin{abstract}
Primordial stars are expected to be very massive and hot, producing 
copious amounts of hard ionizing radiation. The best place to study 
 hard ionizing radiation in the local universe is in very metal-deficient 
Blue Compact Dwarf (BCD) galaxies. We have carried out a MMT  
spectroscopic search for [Ne {\sc v}] $\lambda$3426 (ionization 
potential of 7.1 Ryd), [Fe {\sc v}] $\lambda$4227 
(ionization potential of 4 Ryd) 
and He {\sc ii} $\lambda$4686 (ionization potential of 4 Ryd) 
emission in a sample of 18 BCDs. We have added data 
from previous work and from the Data Release 3 of the Sloan Digital Sky 
Survey. In total, we have assembled a BCD high-ionization sample with 
[Ne {\sc v}] emission in 4 galaxies, [Fe {\sc v}] emission in 15 galaxies and 
He {\sc ii} emission in 465 galaxies. With this large sample, we have reached 
the following conclusions. There is a general trend of 
higher [Ne {\sc v}], [Fe {\sc v}] and He {\sc ii} emission at lower metallicities. 
However metallicity is not the only factor which controls the hardness of 
the radiation. High-mass X-ray binaries and main-sequence stars are 
probably excluded as the main sources of the very hard ionizing radiation
responsible for [Ne {\sc v}] emission. The most likely source of [Ne {\sc v}] 
emission is probably fast radiative shocks moving with velocities 
$\ga$ 450 km s$^{-1}$ through a dense 
interstellar medium with an electron number density of several hundreds per
cm$^{-3}$ and
associated with supernova explosions of the most massive stars. These have 
masses of $\sim$ 50 -- 100 $M_\odot$ and are formed in very compact super-star 
clusters. The softer ionizing radiation required for He {\sc ii} emission is 
likely associated with less massive evolved stars and/or radiative
shocks moving through a less dense interstellar medium.        
\end{abstract}


\section{INTRODUCTION}

Galaxy formation is one of the most fundamental problems in
astrophysics.  To understand how galaxies form, we need to unravel how
stars form from the primordial gas and how the first stars interact
with their surrounding environments.  As there are no heavy elements
in the early universe, the thermodynamic behaviour of the gas is
essentially controlled by H$_{2}$ cooling, and the first Population
III stars are expected to be very massive and hot \citep[e.g.,][]{A02,B02}.
Their effective temperatures are around 10$^5$ K, so they emit 
very hard ionizing radiation with photon energies exceeding
4 Ryd (54.4 eV). These first stars are thus 
very effective in ionizing hydrogen and helium (twice), and 
strong He {\sc ii} emission lines are expected to be present
in the spectra of primeval galaxies \citep[e.g.,][]{S02,S03}. Emission 
lines of high-ionization ions of heavy elements
should be present as well.

While much progress has been made in finding large
populations of galaxies at high ($z\geq3$) redshifts
\citep[e.g.,][]{St96}, truly young galaxies in the process of
forming remain elusive in the distant universe. The spectra of those
far-away galaxies generally indicate the presence of a substantial
amount of heavy elements, implying previous star formation and metal
enrichment. 

Instead of focussing on high-redshift galaxies, another approach is to study 
the properties of the massive stellar populations and their
interaction with the ambient interstellar medium (ISM) in a class of nearby
metal-deficient dwarf galaxies called Blue Compact Dwarf (BCD)
galaxies, some of which are thought to be undergoing their first
episode of star formation \citep{IT04b}. These galaxies have a heavy
element mass fraction $Z$ in the range 1/50 -- 1/3 $Z_\odot$ \citep{IT99},
assuming a solar oxygen abundance 12 + log (O/H)$_\odot$ = 8.91.
Thus, their massive stellar populations have properties intermediate
between those of massive stars in solar-metallicity galaxies such as the 
Milky Way and those of the first stars. BCDs constitute then excellent 
nearby laboratories for studying high-ionization emission and 
the hardness of the ionizing radiation in a very metal-deficient environment.  

The hardness of the ionizing radiation in BCDs has long been known to 
increase with decreasing
metallicity \citep[e.g.,][]{C86}. This is supported by the fact that the 
strong nebular emission line He {\sc ii} $\lambda$4686 is often seen in the 
spectra of BCDs, with a flux which increases with decreasing metallicity
of the ionized gas \citep{GIT00}.
Besides He {\sc ii} emission, high-ionization emission
lines of heavy elements ions are also seen in the spectra of some BCDs. 
 \citet{F01} and \citet{ICS01} first detected the high-ionization emission
line [Fe {\sc v}] $\lambda$4227 in the BCDs
Tol 1214--277 and SBS 0335--052. The presence of this line,
just as that of the He {\sc ii} $\lambda$4686 line, 
requires ionizing radiation with photon energies in excess of 4 Ryd.
More impressively, \citet{ICS01} discovered [Fe {\sc vi}] --
[Fe {\sc vii}] emission in SBS 0335--052 implying that this BCD contains 
intense radiation with photon energies above $\sim$ 7 Ryd ($\sim$ 100 eV), 
i.e. in the range of soft X-rays. 
Note that the ionization potential of
Fe$^{5+}$ is 
5.5 Ryd and that of Fe$^{6+}$ is 
7.3 Ryd.
Later, \citet{I04a} discovered [Ne {\sc v}] $\lambda$3346, 3426 emission lines
in the spectrum of Tol 1214--277. The existence of these lines requires
the presence of hard radiation with photon energies above 7.1 Ryd.
Such hard ionizing radiation is confirmed by the detection of 
[Ne {\sc iv}] (the ionization potential of Ne$^{3+}$ is 
4.7 Ryd) 
and [Fe {\sc vi}] -- [Fe {\sc vii}] emission in a new spectrum of 
Tol 1214--277 \citep{I04c}.

While the presence of hard radiation is well established in some BCDs, the
origin of this radiation is much less clear, in spite 
of several attempts to account for it
\citep[e.g.,][]{G91,SV98}. 

Several mechanisms for producing hard ionizing radiation have been proposed,
such as massive main-sequence stars \citep{S97}, Wolf-Rayet stars \citep{S96a},
primordial zero-metallicity stars \citep{S02,S03}, high-mass X-ray binaries
\citep{G91} and fast radiative shocks \citep{DS96}. However, no mechanism has
emerged clearly as the leading candidate, mainly because of the lack of a
large data base to confront the models with observations. Despite the
importance of understanding the high-ionization phenomenon to interpret the
spectra of primordial star-forming galaxies when these are discovered in the
future, very few observations  
of high-ionization emission lines in metal-deficient BCDs exist. 
The He {\sc ii} $\lambda$4686 emission line has been detected in 
  only a few dozens BCDs, and 
[Ne {\sc v}] and [Fe {\sc v}] -- [Fe {\sc vii}] emission has 
been seen in only two BCDs \citep{GIT00,F01,ICS01,I04a,I04c}. 
This scarcity of data is partly due to the faintness of these high-ionization 
lines (less than 5\% of the flux of H$\beta$ in the case 
of He {\sc ii} $\lambda$4686, and less than 2\% of H$\beta$ in the 
case of the [Ne {\sc v}] $\lambda$3346, 3426 emission lines) and 
their detection requires a 4 m-class telescope or larger. 
To increase the number of known BCDs with
high-ionization emission lines, and construct a reasonably large sample with 
which to study statistical 
trends of high-ionization emission with metallicity and other parameters,
  we have embarked in a program to obtain high signal-to-noise 
spectra in the blue wavelength region of a selected sample of  
BCDs with the 6.5m MMT. To augment our sample, we 
have also carried out a systematic search of the Data Release 3 (DR3) of
the Sloan Digital Sky survey (SDSS) \citep{A05} for 
emission line galaxies with detected He {\sc ii} $\lambda$4686 and
[Fe {\sc v}] $\lambda$4227 emission lines. Finally, for our statistical 
studies, we have also included several more  
BCDs with detected He {\sc ii} $\lambda$4686 emission and three BCDs with 
detected [Fe {\sc v}] $\lambda$4227 emission from the 
sample which \citet{IT04a} used to determine the primordial 
Helium abundance.

We describe the new observational data in Section \ref{Observ}, and how we 
derive element abundances in Section \ref{Abund}. We describe in 
Section \ref{Sample} our sample of 
galaxies with high-ionization emission lines assembled from present and 
previous spectroscopic work. We use that sample for  
statistical studies of 
the high-ionization emission in Section \ref{High}.
We discuss possible mechanisms for 
the hard radiation in Section \ref{Nature}. We explore how the
compactness of the star-forming region may play a role in high-ionization
emission in Section \ref{Compact}, 
and summarize our conclusions in Section \ref{Concl}.

\section{MMT OBSERVATIONS AND DATA REDUCTION \label{Observ}}

We have obtained new high signal-to-noise ratio spectrophotometric 
observations for 26 H {\sc ii} regions in 18 BCDs and for one supergiant 
H {\sc ii} region in the spiral galaxy M101 with
the 6.5-meter MMT on the nights of 2004 February 19 -- 20
and 2005 February 4. In selecting the observational sample, we have been  
guided by several considerations. 
First, we wish the 
equivalent width of the H$\beta$ emission line in the H {\sc ii} regions 
to span a reasonably large range. 
Since EW(H$\beta$) is an indicator of the age of the most recent 
burst of star formation in the BCD, 
this ensures that our BCD sample spans a large burst age range, and 
gives us some handle on the time scale of the high-ionization 
emission phenomenon.   
Second, since the study of how high-ionization emission depends on 
metallicity is one of our main scientific objectives,
we have chosen H {\sc ii} regions
with metallicities spanning the entire metallicity range of BCDs, 
from $\sim$ $Z_\odot$/50 to $\sim$ $Z_\odot$/3.
Third, we have been careful to include BCDs with and without Wolf-Rayet (WR) 
stars since we want to study whether high-ionization emission is 
correlated or not with the presence of WR stars. Fourth, we have included 
objects known from previous spectroscopic work (our 4m or SDSS spectra) to 
have high-ionization emission, He {\sc ii} $\lambda$4686 or 
[Fe {\sc v}] $\lambda$4227 lines, to check if the 
presence of these lines necessarily implies that higher ionization 
lines are present. Finally, we also include BCDs with 
different morphologies. Star formation in BCDs appears to occur in two 
different modes: a relatively quiescent ``passive'' 
mode without super-star cluster (SSC) formation, with a low star formation 
rate (SFR) in a diffuse 
star-forming region, and a considerably more ``active'' mode with SSC 
formation, a  
high SFR in a very compact star-forming region.
This will allow us to check whether the presence of high-ionization 
emission is related to the compactness of the H {\sc ii} region.
 
The 18 BCDs and the H {\sc ii} region in the spiral galaxy M101 
observed with the MMT are listed in Table \ref{Tab1} in order of increasing
right ascension, along with some of their general properties such as 
coordinates, apparent magnitudes, redshifts and absolute magnitudes. 
In addition, we have also included 
in Table \ref{Tab1} BCDs previously observed with other
telescopes which show in their spectra the high-ionization [Fe {\sc v}] 
$\lambda$4227 emission line. Three BCDs were observed by
\citet{IT04a} with the Kitt Peak 4m telescope and five others were selected
from the DR3 of the SDSS \citep{A05}.

   All observations were made with the Blue Channel of the MMT 
spectrograph. We used a 
2$''$$\times$300$''$ slit and a 800 grooves/mm grating in first
order. The above instrumental set-up gave a spatial scale
along the slit of 0\farcs6 pixel$^{-1}$, a scale perpendicular to the slit
of 0.75\AA\ pixel$^{-1}$, a spectral range of 3200--5200\AA\ and a spectral
resolution of $\sim$ 3\AA\ (FWHM). 
The seeing was in the range 1$\arcsec$--2$\arcsec$. 
Total exposure times varied 
between 10 and 105 minutes. Each exposure was broken up into 2--7 
subexposures, not exceeding 15 minutes, to allow for removal of cosmic rays. 
     Three Kitt Peak IRS spectroscopic standard stars, G191B2B, Feige 34 and 
HZ 44 were observed at the beginning, middle and end of each night
for flux calibration. Spectra of He--Ar comparison arcs were obtained
before and after each observation to calibrate the wavelength scale. The log
of the observations is given in Table \ref{Tab2}.

    The two-dimensional spectra were bias subtracted and flat-field corrected
using IRAF\footnote{IRAF is distributed by National Optical Astronomical 
Observatory, which is operated by the Association of Universities for 
Research in Astronomy, Inc., under cooperative agreement with the National 
Science Foundation.}. We then use the IRAF
software routines IDENTIFY, REIDENTIFY, FITCOORD, TRANSFORM to 
perform wavelength
calibration and correct for distortion and tilt for each frame. 
 Night sky subtraction was performed using the routine BACKGROUND. The level of
night sky emission was determined from the closest regions to the galaxy 
that are free of galaxian stellar and nebular line emission,
 as well as of emission from foreground and background sources.
One-dimensional spectra were then extracted from each two-dimensional 
frame using the APALL routine. Before extraction, distinct two-dimensional 
spectra of the same H {\sc ii} region
were carefully aligned using the spatial locations of the brightest part in
each spectrum, so that spectra were extracted at the same positions in all
subexposures. For all objects, we extracted the 
brightest part of the BCD, using a 6\arcsec$\times$2\arcsec\ extraction 
aperture.
In the case of 
five BCDs, Mrk 71, I Zw 18, Mrk 35, Mrk 178 and Mrk 59, 
spectra of 2 -- 3 different H {\sc ii} regions in the same galaxy were 
extracted. 
All extracted spectra from the same object were then co-added. 
We have summed the individual spectra 
from each subexposure after manual removal of the cosmic rays hits. 
The spectra obtained from each subexposure
were also checked for cosmic rays hits at the location of strong 
emission lines, but none was found.

Particular attention was paid to the derivation of the sensitivity curve. 
It was obtained by 
fitting with a high-order polynomial the observed spectral energy 
distribution of the bright hot white dwarf standard stars G191B2B, Feige 34 
and HZ 44. Because the spectra of these stars have only a small number of a 
relatively weak absorption features, their spectral energy distributions are 
known with very good accuracy \citep{O90}. 
Moreover, the response function of the CCD detector is smooth, so we could
derive a sensitivity curve with a precision better than 1\% over the
whole optical range. 

We show in Figure \ref{Fig1}
the spectrum with labeled emission lines of one of the most interesting 
BCDs in our sample, SBS 0335--052E. This BCD with a metallicity of 
only $\sim$ 2.5\% that of the Sun \citep{I97c,I99}, is one of the 
most metal-deficient star-forming galaxies known in the local universe.
It has a SFR of $\sim$ 0.4 $M_\odot$ yr$^{-1}$ and most of the 
star formation occurs in six SSCs within a region of $\sim$ 500 pc in size
\citep{T97}. Several 
high-ionization emission lines are seen in its spectrum. In addition
to the [Fe {\sc v}] $\lambda$4227 emission line already 
discussed by \citet{F01} and \citet{ICS01},
we report here the discovery of the [Ne {\sc iv}] $\lambda$4725 and 
[Ne {\sc v}] $\lambda$3346, 3426 emission lines. 
This makes SBS 0335--052E 
only the second BCD known to have [Ne {\sc iv}] and [Ne {\sc v}] 
emission, after Tol 1214--277 \citep{I04a,I04c}. This discovery 
confirms and strengthens
the finding by \citet{ICS01} who discovered 
[Fe {\sc vi}] -- [Fe {\sc vii}] emission in SBS 0335--052E and 
concluded that ionizing radiation above $\sim$ 7 Ryd must be 
intense in this galaxy. We note that, except for a very 
few papers from our group \citep[e.g.,][]{I04a,I04c},
the 3200--3700 \AA\ spectral region has not been explored extensively 
before for BCDs. Not only does this spectral region 
contain the high-ionization
[Ne {\sc v}] lines discussed before, it also includes the Balmer jump, 
He {\sc i} recombination lines, and in the case of SBS 0335--052E which has a 
higher redshift, the He {\sc ii} $\lambda$3203 line. 

The spectra for the other 26 H {\sc ii} regions observed with
the MMT are shown in Figure \ref{Fig2}. The four dotted vertical lines show
respectively the locations of the [Ne {\sc v}] $\lambda$3346, 3426, 
[Fe {\sc v}]
$\lambda$4227 and He {\sc ii} $\lambda$4686 emission lines.
All spectra in Figs. \ref{Fig1} and 
\ref{Fig2} have been reduced to zero redshift. 
Of note is the detection 
of [O {\sc ii}] $\lambda$3727 and [O {\sc iii}] $\lambda$5007 emission
lines in the spectrum of the C component of the second-most metal-deficient 
star-forming galaxy 
known, I Zw 18, with a metallicity of $\sim$2\% that of the Sun. I Zw 18 C is 
a star-forming complex located $\sim$ 22\arcsec\ northwest of the 
main star-forming region of I Zw 18. 
 Although I Zw 18 C  has been studied spectroscopically before
\citep{IT98,Z98,I01c}, no oxygen lines have been 
detected. The MMT observations 
show conclusively that the ionized gas in the C component of I Zw 18 
does contain heavy elements, just as its main component. 
We also point out the detection of the Bowen fluorescent 
O {\sc iii} $\lambda$3092 emission
line \citep{A84} in the spectrum of the highest-redshift BCD in the sample,
J 0519+0007 (Table \ref{Tab1}). This is the first such detection
in the spectra of a low-metallicity BCD. The presence of this line indicates 
intense Ly$\alpha$ emission in the star-forming region of J 0519+0007.

   The observed line fluxes $F$($\lambda$) normalized to $F$(H$\beta$) 
and multiplied by a factor of 100 and
their errors, for the 27 H {\sc ii} regions shown in Figs. \ref{Fig1} and 
\ref{Fig2} are 
given in Table \ref{Tab3}. They were measured using the IRAF SPLOT routine.
The line flux errors listed include statistical errors derived with
SPLOT from non-flux calibrated spectra, in addition to errors introduced
in the standard star absolute flux calibration.
 Since the differences between the response curves derived for the 
three standard
stars are not greater than 1\%, we set the errors in flux calibration
to 1\% of the
line fluxes. The line flux errors will be later propagated into the 
calculation of abundance errors.
The line fluxes were corrected for both reddening \citep{W58} 
and underlying hydrogen stellar absorption derived simultaneously by an 
iterative procedure as described in \citet{ITL94}. The corrected line 
fluxes 100$\times$$I$($\lambda$)/$I$(H$\beta$), equivalent widths 
EW($\lambda$), extinction
coefficients $C$(H$\beta$), and equivalent widths EW(abs) of the hydrogen
absorption stellar lines are also given in Table \ref{Tab3}, along with the 
uncorrected H$\beta$ fluxes. 

\section{PHYSICAL CONDITIONS AND DETERMINATION OF HEAVY ELEMENT ABUNDANCES
\label{Abund}}

   To determine element abundances, we follow generally
the procedures of \citet{ITL94,ITL97} and \citet{TIL95}. 
We adopt a two-zone photoionized H {\sc ii}
region model: a high-ionization zone with temperature $T_e$(O {\sc iii}), 
where [O {\sc iii}], [Ne {\sc iii}] and [Ar {\sc iv}] lines originate, and a 
low-ionization zone with temperature $T_e$(O {\sc ii}), where [N {\sc ii}], 
[O {\sc ii}], [S {\sc ii}] and [Fe {\sc iii}] lines originate. 
As for the [S {\sc iii}] and [Ar {\sc iii}] lines, they originate in the 
intermediate zone between the high and low-ionization regions \citep{G92}. 
We have derived the chlorine abundance from [Cl {\sc iii}] emission lines in 2 
H {\sc ii} regions. We assume that these lines also originate in the 
intermediate zone as 
the ionization potentials of the S$^{2+}$, Ar$^{2+}$ and Cl$^{2+}$ ions are 
similar. The temperature $T_e$(O {\sc iii}) is calculated using the 
[O {\sc iii}] $\lambda$4363/($\lambda$4959+$\lambda$5007) ratio. To take into 
account the electron temperatures for different ions, we have used 
the H {\sc ii} photoionization models of \citet{SI03}, as fitted by 
the expressions in \citet{I05}. These expressions are based on 
more recent stellar atmosphere models and updated atomic collisional
strengths as compared to those of \citet{ITL94,ITL97}.
Since our observations cover 
only the blue part of the optical spectrum, the [S {\sc ii}] 
$\lambda$6717, 6731 emission lines usually used to 
determine the electron number density
$N_e$(S {\sc ii}) were not available. Therefore, we set 
$N_e$(S {\sc ii}) = 10 cm$^{-3}$. The low-density limit for 
abundance determinations should hold as long as $N_e$ is less than 10$^4$
 cm$^{-3}$. 
Ionic and total heavy element abundances for the 24 H {\sc ii} regions 
observed with the MMT are derived in the manner described 
in \citet{I05} and are given in 
Table \ref{Tab4} along with the adopted ionization correction factors $ICF$.
We do not derive element abundances for I Zw 18C, Mrk 35 (\#2) and 
Mrk 178 (\#3) as the electron temperature-sensitive 
[O {\sc iii}] $\lambda$4363 emission line is not detected in the
spectra of those H {\sc ii} regions. We have compared 
the abundances derived here with previous determinations in several objects.
There is general good agreement, the differences between 
independent measurements not exceeding $\sim$ 0.05 dex. We note that
in SBS 0335--052W \citep[the star-forming dwarf galaxy
which shares the same large H {\sc i} envelope with SBS 0335--052E, 
see][]{P01},
 the oxygen abundance 12 + log O/H = 7.11 $\pm$ 0.05 derived
here is lower than the one 
derived by \citet{L99}.
Remarkably, it is lower than in I Zw 18 (this paper), stealing from the 
latter the qualification of ``most metal-deficient star-forming galaxy known 
in the local universe'' \citep{ITG05}.

\section{A SAMPLE OF BLUE COMPACT DWARF GALAXIES WITH 
HIGH-IONIZATION EMISSION LINES \label{Sample}}

Using present and previous observations, we now construct a sample
which contains as many BCDs with high-ionization 
emission as possible. Out of the 
27 H {\sc ii} regions observed with the MMT, we have detected [Fe {\sc v}]
$\lambda$4227 emission line in 8 H {\sc ii} regions, 
including SBS 0335--052E, 
and [Ne {\sc v}] $\lambda$3346, 3426 in 3 H {\sc ii} regions. 
We have also searched for [Fe {\sc v}] $\lambda$4227 emission  
in spectra obtained earlier for our program on the determination of 
the primordial helium abundance \citep{IT04a}, and in the spectra of the SDSS 
DR3.  The search yielded 3 BCDs from the
\citet{IT04a} sample and 5 BCDs from the SDSS.
To these data we have added Tol 1214--277, known to possess 
both [Ne {\sc v}] $\lambda$3346, 3426 and [Fe {\sc v}] $\lambda$4227 emission
\citep{F01,I04a,I04c}. The emission line fluxes and their 
equivalent widths of the objects selected from \citet{IT04a}
and from the SDSS DR3 are shown in Table \ref{Tab5}. The element abundances
in these H {\sc ii} regions are derived in the same manner as for the objects
observed with the MMT and are given in Table \ref{Tab6}. 
In total,
our sample contains 15 
BCDs with detected [Fe {\sc v}] $\lambda$4227 emission (ionization potential 
of 4 Ryd). Out of those 15, 
only 4 show the higher ionization [Ne {\sc v}] $\lambda$3346, 3426
emission line (ionization potential 
of 7.1 Ryd). This represents a substantial increase of the total number
of known star-forming galaxies with high-ionization emission lines in their
spectra. The spectra of galaxies with [Fe {\sc v}] $\lambda$4227 emission 
(indicated by a dotted vertical line) are shown in the wavelength range 
$\lambda$4150 -- 4770 in Fig. \ref{Fig3}. Those with 
[Ne {\sc v}] $\lambda$3346, 3426 emission (indicated by two dotted vertical 
lines) are shown in the wavelength range 
$\lambda$3150 -- 3650 in Fig. \ref{Fig4}.

We have also conducted a search for galaxies with He {\sc ii}
$\lambda$4686 emission (ionization potential 
of 4 Ryd). The resulting sample is considerably larger: 465 star-forming
galaxies, of which 396 galaxies came from the SDSS DR3 and the 
remaining 69 galaxies from our previous observations and the present 
MMT ones. We have also measured line fluxes and derived heavy element
abundances for all of these 
galaxies in the manner described above. However, because of the large
data set, we do not tabulate them but will simply plot them 
in various figures to detect and study statistical trends.

\section{STATISTICAL PROPERTIES OF THE HIGH-IONIZATION SAMPLE \label{High}}

\subsection{Metallicity dependence}

As mentioned in the Introduction, previous spectroscopic 
studies \citep[e.g.,][]{C86}
have revealed that the hardness of the stellar ionizing radiation in BCDs
increases with decreasing metallicity. This trend implies that nebular
emission lines of ions with high ionization potentials are stronger
in galaxies with lower metallicity. We can check this trend with 
our large He {\sc ii} sample. 
 Fig. \ref{Fig5} shows the He {\sc ii} $\lambda$4686
to H$\beta$ flux ratio as a function of the oxygen abundance in the ionized
gas. The dots show individual data points while the open circles show 
the means of the data points in the 7.1 - 7.6,
 7.6 - 8.1 and 8.1 - 8.6 intervals of 12+ logO/H. The error bars show the 
mean error in each interval.
There is an evident trend with metallicity. At low oxygen abundance,
12 + logO/H $<$ 7.6, 
the mean He {\sc ii} $\lambda$4686 flux is  0.018 $\pm$ 0.011
that of the H$\beta$ emission line.
However at high  oxygen abundance, 
12 + logO/H $\ga$ 7.9, the He {\sc ii} 
$\lambda$4686 emission is weaker on average, 
with the mean flux being only 0.010 $\pm$ 0.006
of the H$\beta$ emission line flux.
There is however not a well-defined relation between  He {\sc ii}
$\lambda$4686 emission and metallicity. The data show 
a large scatter with an upper envelope. 
This implies that metallicity is not the only parameter
which governs He {\sc ii} $\lambda$4686 emission. We can also 
check the dependence of other high-ionization line fluxes on metallicity,
albeit with smaller samples.
Figs. \ref{Fig6} and \ref{Fig7} show respectively [Fe {\sc v}] $\lambda$4227
and [Ne {\sc v}]  $\lambda$3346, 3426 emission as a function of ionized gas 
oxygen abundance. There appears also to be a trend of higher 
[Fe {\sc v}] $\lambda$4227 emission toward lower metallicities,
 with [Fe {\sc v}] $\lambda$4227/H$\beta$ = 0.0066 $\pm$ 0.0038 for
12 + logO/H $<$ 7.6 and [Fe {\sc v}] $\lambda$4227/H$\beta$ = 
0.0036 $\pm$ 0.0021 for 12 + logO/H $\ga$ 7.6. 
 As for [Ne {\sc v}] emission, the sample is too small 
to make definite conclusions. However, Fig. \ref{Fig7} does show that the two
lowest-metallicity BCDs have systematically higher
[Ne {\sc v}] $\lambda$3426 fluxes than the two highest-metallicity
galaxies. Of note is the considerably larger [Ne {\sc v}] flux 
(nearly a factor of 4) 
of the BCD Tol 1214--277 (second point from the left) as compared to that 
of SBS 0335--052E which is more metal-deficient.

\subsection{Hardness ratios}
 
To quantify the hardness of ionizing radiation for low-ionization ions, 
\citet{VP88} have introduced the
parameter $\eta$ for the oxygen and sulfur ionic fractions defined as 
$\eta = 
\frac{{\rm O}^+/{\rm O}^{2+}}{{\rm S}^+/{\rm S}^{2+}}$.
We cannot do the same for our high-ionization ions as we have no
observational constraints on the electron temperatures in the He$^{2+}$,
Ne$^{4+}$ and Fe$^{4+}$ zones of the H {\sc ii} regions. Therefore, we
simply compare the flux ratios of emission lines for different ions of
the same element.
In Table \ref{Tab7} we show these flux ratios for each galaxy 
with detected heavy element high-ionization emission lines,
with the means and dispersions of these ratios for two subsamples, 
a low-metallicity subsample with 12 + logO/H $\leq$ 7.6, and a 
higher-metallicity subsample with 12 + logO/H $>$ 7.6.
Again, there is clear indication
from comparison of these mean values that the hardness of the ionizing 
radiation is higher for the lower-metallicity subsample.
We do not compute the mean values for the 
$I$([Ne {\sc v}] $\lambda$3426)/$I$([Ne {\sc iii}] $\lambda$3868) flux ratio 
because
of poor statistics,
 but it is clear from Table \ref{Tab7} that this ratio for the
two more metal-deficient galaxies SBS 0335--052 and Tol 1214--277 
is significantly 
greater than that for the two more 
metal-rich galaxies, HS 0837+4717 and Mrk 209.
Thus these hardness flux 
ratios support the conclusion reached in section 5.1, that the ionizing 
radiation becomes harder in lower-metallicity BCDs.  

\subsection{Dependence on the age of the starburst}

While it is clear that the hardness of the ionizing
radiation in BCDs decreases with increasing 
metallicity, the large scatter of the He {\sc ii} $\lambda$4686 
fluxes in Fig. \ref{Fig5}, 
of the [Fe {\sc v}] $\lambda$4227 fluxes in Fig. \ref{Fig6} and 
of the [Ne {\sc v}] $\lambda$3426 fluxes in Fig. \ref{Fig7}, tells us that 
metallicity is not the only 
factor which controls their fluxes. We explore here another 
factor which may influence the hardness of the ionizing radiation in BCDs,
the age of the current starburst.

The age of the starburst may play a role as stars go through 
different evolutionary phases which may modify their ionizing fluxes. 
The equivalent width of the H$\beta$ emission line EW(H$\beta$) is a good 
indicator of the starburst age 
\citep[see, e.g.,][]{SV98}. The greater EW(H$\beta$) is, the younger is
the starburst. Stellar evolution models predict \citep{SV98} that
EW(H$\beta$) is $\sim$ 600\AA\ for a zero-age starburst. 
The origin of nebular He {\sc ii} $\lambda$4686 emission in photoionized 
supergiant H {\sc ii} regions has been a subject of debate for many years,
ever since it was realized that the flux of this line is several times larger 
than model predictions for H {\sc ii} regions photoionized by main-sequence 
stars \citep[e.g.,][]{G91,S96a}. 
In Fig. \ref{Fig8} we show the dependence on EW(H$\beta$) of the 
He {\sc ii} $\lambda$4686 emission line flux relative to the  H$\beta$ flux.
The dots show individual data points while the open circles 
show the means of the data points in the  0 - 100 \AA\,
100 - 200 \AA\ and 200 - 300 \AA\ intervals of  EW(H$\beta$). The error bars 
show the mean error in each interval.
There is no clear general trend. The ratio 
He {\sc ii} $\lambda$4686/H$\beta$ is the same, equal to 0.011 $\pm$ 0.007,
 for either the sample of objects with    
EW(H$\beta$) $<$ 100\AA\ or the one containing 
objects with EW(H$\beta$) $\ga$ 
100\AA. Thus, the hardness of the ionizing radiation
with photon energy above 4 Ryd does not appear to depend on starburst
age. 

There are however two features
to notice in Fig. \ref{Fig8} which constrain the nature of the source of
ionizing radiation. First, in very young bursts with EW(H$\beta$)
$\ga$ 300\AA\ corresponding to a starburst age of $\sim$ 3 -- 4 Myr, there is 
a clear absence of strong He {\sc ii} emission. Such ages correspond to 
lifetimes of very massive stars, with masses $\sim$ 100 $M_\odot$. 
This implies that the ionizing radiation
responsible for He {\sc ii} $\lambda$4686 emission probably does not come
from massive stars
in their main-sequence phase, but rather in their post-main-sequence phase. 
Second, strong He {\sc ii} $\lambda$4686
emission is detected in many starbursts with EW(H$\beta$) $\sim$ 50 -- 300\AA,
corresponding to starburst ages in the range $\sim$ 3 - 10 Myr. Such ages 
correspond to lifetimes of main sequence stars in the mass range from 
from 50 to 15 $M_\odot$. 
Thus, it appears that, 
 not only the most massive evolved stars are responsible for
the hard ionizing radiation, but also less massive stars and/or their
descendants and remnants. 

Next, we explore other
possible sources of hard ionizing radiation in low-metallicity BCDs.

\section{SOURCES OF HARD IONIZING RADIATION \label{Nature}}

\subsection{Wolf-Rayet stars}

If present photoinization models of main-sequence stars cannot
account for the intensity of nebular He {\sc ii} $\lambda$4686 emission,
models for post-main sequence Wolf-Rayet stars also fail. 
\citet{S96a} has 
synthesized the nebular and Wolf-Rayet He {\sc ii} $\lambda$4686 
emission in young starbursts. For heavy element mass fractions $Z_\odot$/5 
$\leq$ $Z$ $\leq$ $Z_\odot$, he predicts strong nebular 
He {\sc ii} emission due to a significant fraction of WC stars 
in the early WR 
phases of the burst. However, \citet{GIT00} have found that the strength of
the He {\sc ii} $\lambda$4686 does not correlate with the 
WR stellar population in the galaxy, so that WR stars cannot be the 
main contributor to He {\sc ii} $\lambda$4686 emission.

\subsection{The hottest normal main-sequence stars and primordial stars} 

\citet{I04a,I04c} have modeled the physical conditions in the BCD 
Tol 1214--277 to reproduce the observed fluxes of the 
He {\sc ii} $\lambda$4686,
[Ne {\sc v}] $\lambda$3426 and [Fe {\sc v}] $\lambda$4227 emission lines.
This BCD is characterized by a metallicity of about 4\% that of the Sun and 
by the strongest [Ne~{\sc v}] $\lambda$3426\AA\  emission line known.  
To produce strong [Ne~{\sc v}] $\lambda$3426\AA, 
the ionizing radiation must be intense at $\lambda$ $\la$ 128\AA\ because
the ionization potential of the Ne$^{4+}$ ion is 
7.1 Rydberg.
Using Kurucz and CoStar stellar atmosphere models, 
\citet{I04a,I04c} have calculated several spherically symmetric 
ionization-bounded H {\sc ii} region models which 
reproduce reasonably well the observed emission line fluxes of the O$^+$, 
O$^{2+}$, Ne$^{2+}$, S$^{2+}$, Ar$^{2+}$ and Ar$^{3+}$ ions. However,
the models fail to reproduce 
the observed [Ne {\sc v}] $\lambda$3426/[Ne {\sc iii}] $\lambda$3868 
 and He {\sc ii} $\lambda$4686/H$\beta$ 
flux ratios in Tol 1214--277, 
the predicted values being respectively
some 10$^4$ and 50 times smaller respectively, even with 
the CoStar atmosphere models of the hottest main-sequence stars from 
\citet{S97} (their F1 model with $T_{\rm eff}$ $\approx$ 54000 K).
The models also fail to reproduce the high-ionization line ratios 
in less extreme objects such as Mrk 209.  
The F1 model predicts 
[Ne {\sc v}] $\lambda$3426/[Ne {\sc iii}] $\lambda$3868 
 and He {\sc ii} $\lambda$4686/H$\beta$ line flux ratios that are 
smaller than the observed ones by factors of 
10$^3$ and 10 respectively.  
The difference between the observations and
model predictions is even larger when \citet{K79} stellar 
atmosphere models are used.

While the gas in our BCDs is not as metal-deficient as the primordial gas,
we have also compared our observations with 
high-ionization line fluxes predicted for primordial stars 
\citep{S02,S03}. We found that   
it is possible, in principle, to reproduce the 
observed He {\sc ii} $\lambda$4686, [Fe {\sc v}] $\lambda$4227
and [Ne {\sc v}] $\lambda$3426 emission line fluxes by models of  
 very low metallicity ($Z$ $\la$ 10$^{-7}$)  massive
ionizing stars. However, such models of Population III stars 
would predict equivalent widths of
H$\beta$ and Ly$\alpha$ emission lines that are several times
larger than those observed (Tables \ref{Tab3} and \ref{Tab5}). 
Thus 
neither models of normal stars, of Wolf-Rayet stars nor of primordial 
stars are able to reproduce the observed high-ionization line fluxes.

\subsection{High-mass X-ray binaries}

The energy of the 
photons that can produce Ne$^{4+}$ ions are in the soft X-ray range.
Thus high-mass X-ray binaries (HMXBs) may also be an important source of 
ionizing radiation. 
If stellar atmosphere models of normal hot stars fail to provide the 
necessary ionizing radiation, can HMXBs do better? 
\citet{I04a,I04c} have estimated that the X-ray luminosity 
required to reproduce 
the large observed flux of the [Ne {\sc v}] emission in  Tol 1214--277
at wavelengths shorter $\sim$ 100\AA\ (corresponding to a photon energy 
of 0.14 keV) should be
as high as $L_{X}$ = 10$^{39}$ -- 10$^{40}$ erg s$^{-1}$.  
Tol 1214--277 has not been observed in the X-ray range. In fact, 
of the four BCDs 
known to possess [Ne {\sc v}] $\lambda$3426 emission (Fig. 4), 
only SBS 0335--052E 
has {\it Chandra} X-ray observations \citep{T04}. It is found that more 
than 90\% of its 0.5 -- 10 keV 
(or 1 -- 25\AA) flux comes from a point-source with a luminosity 
of 3.5 $\times$ 10$^{39}$ erg s$^{-1}$. 
If that point source is composed of a single object,
then its luminosity would place it in the range of 
the so-called ultraluminous X-ray sources (ULXs). 

\citet{T04} have suggested that the high X-ray luminosities 
of these sources may be due to a metallicity effect. 
The X-ray heating of the gas as it falls toward a compact object depends 
strongly on the atomic number Z, since the photoelectric cross-section varies 
as Z$^4$. Thus in low-metallicity systems, the X-ray heating is less, 
resulting in a larger accretion rate and X-ray luminosity \citep{PM95}.
It is also plausible that a lower 
gaseous metallicity may also 
result in higher mass accreting black holes, first 
by allowing the formation of more massive progenitor stars because of 
reduced cooling, and second 
by helping the formation of a more massive compact object from a normal 
progenitor OB star. A lower metallicity results in the reduction of the 
mass-loss rate in the radiatively driven winds, leading to a more massive 
stellar core, which may then collapse into a more massive black hole. 
 
The X-ray spectrum 
of the point source in SBS 0335--052E is 
well fitted by a moderately soft power law \citep{T04},
typical of X-ray emission from HMXBs. 
The observed [Ne {\sc v}] $\lambda$3426/[Ne {\sc iii}] $\lambda$3868 
flux ratio in SBS 0335--052E is $\sim$ 2.7 times smaller than that in 
Tol1214--277 (Table 7), so that the X-ray luminosity of the ULX observed 
in it \citep{T04}, can in principle account for the [Ne {\sc v}] emission.  
 However, based on the very scarce X-ray data on low-metallicity BCDs, 
there does not appear to be a one-to-one correlation between the 
presence of a ULX and [Ne {\sc v}] emission.   
Thus the [Ne {\sc v}] $\lambda$3426 emission line is not seen in the 
MMT spectra (Fig. \ref{Fig2}) of the BCDs I Zw 18 with a metallicity of 
about 2\% that of the 
Sun, and Mrk 59 with a metallicity of about 12\% that of the Sun, although 
a X-ray point-source has been detected in I Zw 18 \citep{T04}, 
and two in Mrk 59 (Thuan et al. 2005, in preparation),   
with X-ray luminosities comparable to the one in SBS 0335--052E \citep{T04}.
This lack of correlation appears to rule out HMXBs as mainly responsible 
for the high-ionization [Ne {\sc v}] $\lambda$3426 emission.

\subsection{Fast radiative shocks}

\citet{G91} have suggested that fast radiative 
shocks in giant H {\sc ii} regions can produce relatively strong He {\sc ii} 
emission under certain conditions.
Hydrodynamical models by \citet{DS96} have indeed shown that  
fast shocks are an efficient means to generate 
a strong local UV photon field in the ISM of a galaxy.
The total optical and UV emission from the shock and the precursor region 
scales as the mechanical energy flux through the shock.
The H$\beta$ luminosity generated in the shock region which measures 
the number of recombinations that occur in the ionized gas column behind the 
shock provides a convenient way for estimating the total 
mechanical energy dissipation rate through the shock structure.   
With shock velocities of $\sim$ 400 -- 500
km s$^{-1}$, examination of the \citet{DS96} models shows that 
if the H$\beta$ luminosity generated in the shock region is
several percent of the H$\beta$ luminosity produced by massive stars, such
shocks can be responsible for the observed fluxes of both 
the He {\sc ii} $\lambda$4686\AA\
and [Ne {\sc v}] $\lambda$3346, 3426\AA\ emission lines
for all four BCDs with [Ne {\sc v}] $\lambda$3346, 3426 emission, 
the spectra of which are shown in Fig. \ref{Fig4}. 

There is an observational check for the presence of such fast 
radiative shocks. 
Their large velocities should produce broad components in the line profiles of 
the strong emission lines. Indeed, broad wings of the H$\beta$ and 
[O {\sc iii}] $\lambda$4959, 5007 emission lines are seen in the spectra
of all four BCDs with detected [Ne {\sc v}] $\lambda$3426 emission:
SBS 0335--052 (Fig. \ref{Fig1}), HS 0837+4717 and Mrk 209 (Fig. \ref{Fig2})
and Tol 1214--277 \citep[Fig. 2 in ][]{I04c}, and the broad component fluxes
are $\sim$ 1\% -- 2\% of the narrow component fluxes. 
The full widths at half maximum FWHM $\ga$ 15 \AA\ of 
the broad components
correspond to gas velocities in expanding shells of $\ga$ 900/2 km s$^{-1}$
= 450 km s$^{-1}$.
The [Ne {\sc v}] emission and the broad components in the 4 BCDs appear to be
related to the evolution of the most massive stars. This is because
their H$\beta$ equivalent widths are high, 
EW(H$\beta$) $\ga$ 190\AA\ (Table \ref{Tab3}), corresponding to a starburst
age $\la$ 3 -- 4 Myr, the main-sequence lifetime of a $\ga$ 50 -- 100
$M_\odot$ star. Thus, on the basis 
of a meager sample of four objects, [Ne {\sc v}] $\lambda$3426 emission 
does appear to be accompanied by fast gas velocities 
of $\sim$ 450 km s$^{-1}$, in the range of velocities expected to be 
produced by fast 
shocks. However, the reverse is not true: the presence of broad components 
is not necessarily accompanied by high-ionization emission. For example,
broad wings of the strong emission lines are also seen in the spectra
of II Zw 40, Mrk 71 and Mrk 59 (Fig. \ref{Fig2}). However, their spectra 
which have a comparable signal-to-noise ratio, do 
not show [Ne {\sc v}] $\lambda$3426 emission. Metallicity probably 
plays a role here. The latter three galaxies with broad
wings of the strong emission lines but without [Ne {\sc v}] emission are
all more metal-rich (their 12 + log O/H are respectively 8.1, 7.9 and 
8.0) than the four BCDs with broad wings and 
[Ne {\sc v}] emission (their 12 + log O/H are respectively 7.3, 7.6, 
7.8 and 7.5). It is probable that in higher-metallicity BCDs, for a given 
shock velocity, the shock-heated ionized gas cools more efficiently and hence 
the high-ionization postshock region is narrower, so that 
the fraction of  
highly ionized ions is lower than that in lower-metallicity BCDs.
 Unfortunately, we cannot check this metallicity dependence 
directly with hydrodynamical calculations as 
\citet{DS96} computed only models with solar metallicity. 

In summary, existing models of massive main-sequence stars and of
post-main-sequence Wolf-Rayet stars are unable to account for the hard
radiation responsible for He {\sc ii} $\lambda$4686 emission, let alone
for the considerably harder radiation responsible for 
[Ne {\sc v}] $\lambda$3426 emission. HMXBs may play a role although the scarce
X-ray data on high-ionization BCDs show that there is not a one-to-one
relationship between X-ray and [Ne {\sc v}] $\lambda$3426 emission. Fast
radiative shocks with velocities  $\sim$ 450 km s$^{-1}$
associated with the evolution of massive 
($M$ $\sim$ 50 -- 100 $M_\odot$) stars appear to be the best candidate in town.
The presence of such
fast motions is supported by the broad components observed in the line 
profiles of the strong emission lines. Fast radiative shocks 
are also able
to account for the general trend of more intense high-ionization emission at
lower metallicities, since the high-ionization postshock region should become
wider with decreasing heavy element abundances and less cooling.

\section{HIGH-IONIZATION EMISSION AND SUPER-STAR CLUSTERS 
IN BLUE COMPACT DWARF GALAXIES \label{Compact}}

We discuss in this section how the compactness of the star-forming regions in
BCDs may play an important role in the high-ionization phenomenon.

\subsection{Imaging data and super-star clusters}

One striking common feature which links the BCDs with detected 
[Ne {\sc v}] $\lambda$3426 emission in Fig. \ref{Fig4} is the presence of
luminous, compact and high-surface-brightness Super-Star Clusters (SSCs). 
This can be seen in Fig. \ref{Fig9} which shows the {\sl HST} images
of SBS 0335--052E, Tol 1214--277 and Mrk 209, and the SDSS image 
of HS 0837+4717. We will use broad-band images to discuss the sizes and 
luminosities of the SSCs. 

The Tol 1214--277 image was obtained by us using the WFPC2 camera aboard  
{\sl HST}\footnote{Based on observations 
obtained with the NASA/ESA {\sl Hubble Space Telescope} through the Space 
Telescope Science Institute, which is operated by AURA,Inc. under NASA
contract NAS5-26555.} on 1996 June 29 (PI: T. X. Thuan, proposal ID No. 6678).
In addition to the $V$ (F555W filter, 1600 s exposure time) 
image shown in Fig. \ref{Fig9},
we have also obtained an $I$ (F814W filter, 3200 s exposure time) image.
Tol 1214--277 was centered on the WF3 chip, giving an 
angular resolution of 0\farcs101 per pixel. A 
pixel corresponds to a spatial extent of
50 pc at the distance of 103 Mpc of the BCD.
Fig. \ref{Fig9}
shows that star formation in 
Tol 1214--277 occurs primarily in a bright compact stellar
cluster. There is a 
low-surface-brightness component to the south-west of the 
cluster. The compact cluster has a FWHM of 3.8 pixels, 
corresponding to a spatial
extent of 190 pc. Its $V$, $I$ magnitudes and $V-I$ color are respectively 
equal to
18.62 $\pm$ 0.01 mag, 19.42 $\pm$ 0.01 mag and --1.20 $\pm$ 0.02 mag. At the 
distance of Tol 1214--277, this corresponds to $M(V)$ = --16.44 mag. 
If this luminosity comes from a single object, then the brightness of 
the SSC in Tol 1214--277 would be more than 5 times brighter than the 
brightest SSC in      
SBS 0335--052E \citep{T97}. However, because 
Tol 1214--277 is nearly 2 times more 
distant than SBS 0335--052E, the linear resolution is poor and we cannot 
exclude the possibility that the star-forming region is composed  
of several SSCs. We have checked its surface brightness profile and find 
it to be
symmetric. Thus either the star-forming region in Tol 1214--277 contains a
single SSC, or if it contains several SSCs, these must be very close to
each other. We note finally that the star-forming region of Tol 1214--277 is 
extremely blue in $V-I$, not only because of the young OB stars contained in 
it, but also because of the strong [O {\sc iii}] $\lambda$5007 nebular emission
line which, at the redshift $z$ = 0.0260 of Tol 1214--277, 
falls into the $V$ band.

Images of the three other BCDs were extracted 
from the {\sl HST} and SDSS archives.
The UV image of SBS 0335--052E 
was obtained by \citet{K03} with the {\sl HST}/ACS camera in 
the F140LP filter (PI: D. Kunth, proposal ID No. 9470).
 The angular resolution is 0\farcs032 per pixel which 
corresponds to a spatial scale of 8.9 pc per pixel at the distance of 
54.3 Mpc. We measured the FWHMs of 
SSCs 1, 4 and 5 \citep[using the notation of ][]{T97} 
to be respectively 5.7, 4.8 and 4.3 pixels, 
corresponding to linear sizes of 51, 43 and 38 pc.
The $J$ image 
of Mrk 209 in the F110W filter was obtained by \citet{S01} 
with the {\sl HST}/NICMOS camera
(PI: R. E. Schulte-Ladbeck, proposal ID No. 7859). 
 The angular resolution is 0\farcs076 per pixel, 
corresponding to a spatial scale of 2.2 pc per pixel
 at the distance of 5.8 Mpc. 
Four bright compact clusters are seen which each have a 
FWHM $\sim$ of 2.2 pixels, 
corresponding to a linear size 4.8 pc. Their apparent $J$ magnitudes are in
the range 18.5 -- 18.7 mag, corresponding to absolute magnitudes $M(J)$ =
--10.3 - --10.1 mag. Since $V - J$ = --0.06 \citep{T83}, 
 $M(V)$ = --10.4 - --10.2. Both these linear sizes and luminosities
put the star clusters in Mrk 209 in the category of SSCs.
Finally, because HS 0837+4717 has not been imaged by {\sl HST}, 
we show in Fig. \ref{Fig9} its $g$-band SDSS image.
Here, the angular resolution is 0\farcs398 per pixel which at the
distance of 168 Mpc corresponds to a spatial scale of 
325 pc per pixel. 
The star-forming region has a size FWHM = 3.9 pixels, corresponding to
a linear size of 1267 pc.
Its apparent $g$ magnitude measured in a 3\arcsec\ aperture is
17.9 mag, corresponding to an 
absolute magnitude --18.2 mag. There are probably several compact clusters
in the star-forming region of HS 0837+4717 which are not resolved 
because of the large
distance of the BCD and the 
poor angular resolution of the SDSS images.
 
Our size and luminosity estimates of SSCs above 
should be considered as upper limits because of the presence of the nebular 
emission which is more extended than the ionizing star clusters.
In principle, both continuum and line nebular 
emission ought to be subtracted from the broad-band images. However, this 
cannot be done as we do not have high spatial resolution narrow-filter images
of the BCDs centered on the emission lines. 
At least for SBS 0335--052E, 
 the overestimate in the SSC sizes is small  
as the contribution of the nebular emission in the UV range at $\sim$
1400\AA\ is minimal.

\subsection{Active vs. passive star formation}

The previous discussion shows that of the four galaxies with detected 
[Ne {\sc v}] $\lambda$3346, 3426 emission, at least three, SBS 0335--052E, 
Tol 1214--277 and Mrk 209, have star formation occurring in very luminous and
compact SSCs. We cannot tell for HS 0837+4717 because it has not been imaged
by {\sl HST}. The compactness of the star-forming region thus appears to  
play a key role in the production of high-ionization radiation with photon 
energy above 7.1 Ryd.

Based on the detailed study of two of the most metal-deficient BCDs known,
I Zw 18 and SBS 0335--052E,
star formation in BCDs appears to occur in two quite different modes: 
SBS 0335--052E makes stars in an ``active'' mode characterized by 
SSC formation, 
a high star formation rate (SFR) of $\sim$ 0.4 $M_\odot$ yr$^{-1}$ \citep{T97},
 a compact size,
hot dust, and significant amounts of H$_2$, while I Zw 18 is characterized by  
a more quiescent ``passive'' mode with an absence of 
SSCs, a low SFR [some 8 times less than 
in SBS 0335--052E, \citet{K94}], a larger size with cooler dust 
and no significant amount of H$_2$. For the purpose of our discussion, 
we will call ``active'' BCDs those with SSCs, and ``passive'' BCDs 
those without. 
 Clearly, metallicity 
is not the distinguishing factor between SBS 0335--052E and I Zw 18 
since both BCDs have similar
heavy element abundance. Theoretical models \citep{H04}
suggest that the difference between the two modes can be 
understood through a difference in size and density 
of the star-forming regions. The active mode occurs in 
regions which are compact (with radius $\la$ 50 pc) and dense (with gas 
number density $\ga$ 500 cm$^{-3}$) as in SBS 0335--052E. 
On the other 
hand, the passive mode occurs in regions that are diffuse (with radius 
$\ga$ 100 pc) and less dense (with gas 
number density $\la$ 100 cm$^{-3}$) as in I Zw 18.

Our preferred mechanism for the source of hard ionizing radiation, fast
radiative shocks, would work best in dense environments with electron
number densities of several hundreds per cm$^{-3}$, similar to
those in ``active'' low-metallicity BCDs with SSC formation.
The luminosities in emission lines scale as the square of the electron number
density. Therefore, shocks moving through a dense ISM
can radiate more and produce high [Ne {\sc v}] emission-line 
luminosities. The electron number densities $N_e$(S {\sc ii}) are indeed 
high in at
least three BCDs with [Ne {\sc v}] emission: $N_e$(S {\sc ii})
= 390 cm$^{-3}$ for SBS 0335--052 \citep{I97c}, 400 cm$^{-3}$ for Tol 1214--277
\citep{ICG01} and 300 -- 400 cm$^{-3}$ for HS 0837+4717 \citep{P04}. 
As for Mrk 209 where [Ne {\sc v}] emission is tentatively detected, its
electron number density is lower, $N_e$(S {\sc ii}) $\la$ 100 cm$^{-3}$
\citep{ITL97}.
On the other hand,  the emission-line luminosities of 
postshock regions in the low-density ISM of ``passive'' BCDs without SSC 
formation are expected 
to be considerably lower, and any [Ne {\sc v}] emission would go undetected.

However, if the presence of fast shocks and dense stellar clusters is a
necessary condition for [Ne {\sc v}] emission, it is not a sufficient
condition. [Ne {\sc v}] emission is absent
in higher-metallicity BCDs, such as II Zw 40, Mrk 59 and Mrk 71, despite
the presence of compact stellar clusters (as examination of archival {\sl HST} 
images shows), and of broad components of the strong emission lines 
(Fig. \ref{Fig2}). A low-metallicity ISM is also needed. Thus, 
the presence of [Ne {\sc v}] emission 
requires the following three conditions to be fulfilled
simultaneously: 1) a low gaseous metallicity; 2) the presence of fast 
shocks and 3) a dense ISM and compact stellar clusters.

\subsection{The decoupling of the very hard and moderate ionizing radiations}

It thus appears that the very high-ionization radiation with photon energy 
greater than 7.1 Ryd is associated with very dense
star-forming regions. To check for the compactness of the [Ne {\sc v}] 
$\lambda$3426 emission, we have plotted in Fig. \ref{Fig10} its 
spatial distribution in the BCD SBS 0335--052E
along the slit, with a position angle of --30 degrees. 
For comparison, we have also plotted the spatial distributions of 
other emission lines, [Fe {\sc v}] $\lambda$4227,  
 He {\sc ii} $\lambda$4686,  H$\beta$ $\lambda$4861, [O {\sc iii}] 
$\lambda$4363 and [O {\sc iii}] $\lambda$5007.
These distributions are similar to those obtained by \citet{ICS01}.
In particular, we confirm the excess of He {\sc ii} $\lambda$4686 emission
in the 
shell at an angular distance of 5\farcs5 to the northwest 
of SSC 1 [see Fig. 10 of \citet{T97}]. 
This excess emission at a location where there is 
no visible star
cluster supports the hypothesis that He {\sc ii} emission 
in SBS 0335--052E is due mainly to radiative shocks.
It is seen from Fig. \ref{Fig10} that the spatial distribution of 
[Ne {\sc v}] $\lambda$3426 emission is more compact than 
those of other high-ionization ions. Thus, 
although the maxima of both [Ne {\sc v}] $\lambda$3426 and 
He {\sc ii} $\lambda$4686 emissions are between SSCs 1,2 and 
SSCs 4,5 [the SSC notation of \citet{T97} is used],
the spatial distribution of the [Ne {\sc v}] emission is not as extended as 
that of the He {\sc ii} emission.  This observation has two consequences.
First, it supports our previous conclusion 
that very hard ionizing radiation 
is associated with the active mode of star formation in BCDs with 
very compact star-forming regions.
Second, it implies that the
source of very hard radiation (with photon energy $\ga$ 7 Ryd) 
responsible for the [Ne {\sc v}] emission is distinct spatially 
from the source of moderate ionizing radiation (with photon energy 
between 4 and 7 Ryd) responsible for 
the He {\sc ii} and [Fe {\sc v}] emission (Fig. \ref{Fig10}a). 
Comparison of the high-ionization emission of two of the most-metal 
deficient BCDs known in the local universe, I Zw 18 and SBS 0335--052E,
supports that contention.
Despite very similar strengths
of the He {\sc ii} $\lambda$4686 emission line 
relative to H$\beta$ in the two BCDs, strong
[Ne {\sc v}] $\lambda$3346,3426 emission is seen only  
in SBS 0335--052E but not in I Zw 18. 
Their nearly equal metallicities
suggest once again that metallicity is not 
the only factor which controls the amount of 
hard ionizing radiation relative to moderate ionizing radiation.     

 Fig. \ref{Fig10}b also shows
that He {\sc ii} $\lambda$4686 emission does not follow
[O {\sc iii}] $\lambda$5007 emission. While the latter is strongest 
at the location of SSC 1, the former is offset in the direction 
of SSCs 4 and 5. This feature has been noted before by \citet{I97c,ICS01}. 
This is consistent with the idea 
that the extended He {\sc ii} $\lambda$4686 emission is due mainly to the 
ionizing radiation from radiative shocks,
in contrast to the [O {\sc iii}] $\lambda$5007 emission which 
originates mainly from the ionizing radiation of main-sequence stars.

\section{SUMMARY AND CONCLUSIONS \label{Concl}}

The first Population III stars are expected to be very massive and hot, 
producing copious amounts of hard ionizing radiation. The best place to 
study hard ionizing radiation in the local universe is in metal-deficient 
blue compact dwarf (BCD) galaxies as the hardness of the ionizing 
radiation in BCDs has long been known to increase with decreasing metallicity.
We have carried out with the 6.5m MMT 
a spectroscopic search for the high-ionization  
[Ne {\sc v}] $\lambda$3426 emission line (ionization potential of 
7.1 Ryd) in 18 BCDs. We have detected [Ne {\sc v}] emission in 
2 BCDs, SBS 0335--052E and HS 0837+4717 and tentatively in Mrk 209. With 
the previous detection of that line in Tol 1214--277 \citep{I04a},
there are now 4 BCDs known to possess [Ne {\sc v}] emission. We have also 
examined the BCD 
spectra for other high ionization lines such as  [Fe {\sc v}] $\lambda$4227 
(ionization potential of 4 Ryd) and  
 He {\sc ii} $\lambda$4686 (ionization potential of 4 Ryd). In order to  
construct a large sample of BCDs with high-ionization lines to study 
statistical trends, we have combined  
the present MMT observations of BCDs with previous ones obtained for our 
primordial helium program \citep{IT04a} and those available in 
the Data Release 3 of the Sloan Digital Sky Survey. We have thus assembled 
15 BCDs with [Fe {\sc v}] $\lambda$4227 emission and 465 BCDs with 
 He {\sc ii} $\lambda$4686 emission. Studies of statistical trends in the 
resulting sample yield the following main results:

1) There is a general tendency for higher He {\sc ii}, [Fe {\sc v}] and 
  [Ne {\sc v}] emission at lower metallicities.  The hardness of the 
radiation as measured by the flux ratios of emission lines for different 
ions of the same element also increases with decreasing metallicity. 
However the scatter is 
large indicating that metallicity is not the only parameter which 
controls the hardness of the ionizing radiation.

2) The hardness of the ionizing radiation with photon energy 
 greater than 4 Ryd does not depend on the burst age as measured by the 
equivalent width of H$\beta$, EW(H$\beta$). There is a clear 
absence of He {\sc ii} emission in very young 3 -- 4 Myr starbursts 
with EW(H$\beta$) $\ga$ 300 \AA, implying that the source of 
ionization is not massive stars in their main-sequence phase, but in their
post-main-sequence phase. 
Strong He {\sc ii} emission 
is seen in starbursts with  EW(H$\beta$) $\sim$ 50 -- 300 \AA, suggesting 
that less massive stars and their descendants also contribute radiation 
to ionize helium. Present photoionization models of main-sequence and 
of Wolf-Rayet stars cannot account for the observed 
He {\sc ii} emission. 

3) The X-ray luminosity in luminous high-mass 
X-ray binaries (HMXBs) observed  in a few very metal-deficient BCDs      
 can account for their [Ne {\sc v}] emission. However, the scarce X-ray data 
does not show a one-to-one correlation between the presence of a HMXB and 
[Ne {\sc v}] emission, so that HMXBs are probably ruled out as the main
source of hard ionizing radiation with energy above 7.1 Ryd.

4) The most likely source of [Ne {\sc v}] emission is probably
fast radiative shocks moving with velocities $\sim$ 450 km s$^{-1}$ 
through a dense ISM with electron number
densities of several hundreds cm$^{-3}$. These 
shocks are probably produced via
the evolution of massive stars with masses of $\sim$ 50 -- 100 $M_\odot$,
formed in very compact and dense super-star clusters (SSCs). 
These fast radiative shocks are 
evidenced by broad components in the line profiles of the 
strong emission lines. However the presence of compact SSCs and 
broad components is not necessarily accompanied by high-ionization emission. 
Metallicity appears to play an important role. [Ne {\sc v}]
emission is detected only in low-metallicity galaxies with
12 + log O/H $\la$ 7.8. In higher-metallicity galaxies
with higher metallicity the postshock regions are cooled more efficiently.
Therefore, their high-ionization regions are
smaller and their emission lower.

5) The spatial distribution of [Ne {\sc v}] emission is more compact 
than that of He {\sc ii} emission, suggesting spatially distinct sources of 
ionizing radiation for the two. While [Ne {\sc v}] emission requires 
hard radiation  produced by fast radiative shocks moving
through a dense interstellar medium, and probably associated with 
the evolution of the most massive stars,  
softer ionizing radiation, likely associated with the evolution of less 
massive stars and/or radiative shocks moving through a lower density
ISM, is required for He {\sc ii} emission. 

\acknowledgments
The MMT time was available thanks to a grant from the 
Frank Levinson Fund of the Peninsula Community Foundation 
to the Astronomy Department of the University of Virginia .
The research described in this publication was made possible in part by Award
No. UP1-2551-KV-03 of the U.S. Civilian Research \& Development Foundation 
for the Independent States of the Former Soviet Union (CRDF).
It has also been supported by NSF grant AST-02-05785.
Y.I.I. thanks the hospitality of the Astronomy Department of 
the University of Virginia. 
All the authors acknowledge the work of the Sloan Digital Sky 
Survey (SDSS) team. 
Funding for the Sloan Digital Sky Survey (SDSS) has been provided by the 
Alfred P. Sloan Foundation, the Participating Institutions, the National 
Aeronautics and Space Administration, the National Science Foundation, the 
U.S. Department of Energy, the Japanese Monbukagakusho, and the Max Planck 
Society. The SDSS Web site is http://www.sdss.org/.
    The SDSS is managed by the Astrophysical Research Consortium (ARC) for 
the Participating Institutions. The Participating Institutions are The 
University of Chicago, Fermilab, the Institute for Advanced Study, the Japan 
Participation Group, The Johns Hopkins University, the Korean Scientist Group, 
Los Alamos National Laboratory, the Max-Planck-Institute for Astronomy (MPIA), 
the Max-Planck-Institute for Astrophysics (MPA), New Mexico State University, 
University of Pittsburgh, University of Portsmouth, Princeton University, the 
United States Naval Observatory, and the University of Washington.


%
\begin{deluxetable}{lccccclc}
\small
\tablenum{1}
\tablecolumns{7}
\tablewidth{0pt}
\tablecaption{General Parameters of Galaxies}
\tablehead{
\colhead{} & \multicolumn{2}{c}{Coordinates (2000.0)} &\colhead{} 
&\multicolumn{2}{c}{ } &\colhead{} \\ 
\cline{2-3} 
\colhead{Name}&\colhead{$\alpha$}&\colhead{$\delta$}&\colhead{$m_{pg}$}&
\colhead{$z$}
&\colhead{$M_{pg}$} &\colhead{Other names}
  }
\startdata
\multicolumn{7}{c}{a) Galaxies observed with the MMT} \\
SBS $0335-052$W & 03$^h$37$^m$38\fs4~        &\,\,\,$-05$\arcdeg02\arcmin37\arcsec&19.0 &0.01367&--14.7&          \\
SBS $0335-052$E & 03\ \ 37\ \ 44.0~        &$-05$\ 02\ 39                     &16.3   &0.01347&--17.4&          \\
J $0519+0007$  & 05\ \ 19\ \ 02.7~        &$+00$\ 07\ 29                      &18.2   &0.04460&--18.1&          \\
II Zw 40       & 05\ \ 55\ \ 42.6~        &$+03$\ 23\ 32                      &15.0   &0.00265&--15.1&UGCA 116  \\
Mrk 71         & 07\ \ 28\ \ 42.5~        &$+69$\ 11\ 21                      &11.0   &0.00031&--14.5&NGC 2363  \\
HS $0822+3542$ & 08\ \ 25\ \ 55.4~        &$+35$\ 32\ 32                      &18.0   &0.00244&--11.9&J $0825+3532$ \\
Mrk 94         & 08\ \ 37\ \ 43.5~        &$+51$\ 38\ 31                      &16.0   &0.00249&--14.0&SBS $0834+518$ \\
HS $0837+4717$ & 08\ \ 40\ \ 29.9~        &$+47$\ 07\ 09                      &18.0   &0.04212&--18.1&J $0840+4707$ \\
SBS $0911+472$ & 09\ \ 14\ \ 34.9~        &$+47$\ 02\ 07                      &16.0   &0.02729&--19.2&          \\
SBS $0926+606$A& 09\ \ 30\ \ 06.4~        &$+60$\ 26\ 53                      &17.0   &0.01370&--16.7&          \\
I Zw 18        & 09\ \ 34\ \ 02.1~        &$+55$\ 14\ 25                      &16.0   &0.00243&--13.9&Mrk 116   \\
SBS $0940+544$ & 09\ \ 44\ \ 16.6~        &$+54$\ 11\ 33                      &18.0   &0.00541&--13.7&          \\
SBS $1030+583$ & 10\ \ 34\ \ 10.2~        &$+58$\ 03\ 49                      &16.0   &0.00753&--16.4&Mrk 1434  \\
Mrk 35         & 10\ \ 45\ \ 22.4~        &$+55$\ 57\ 37                      &13.0   &0.00331&--17.6&Haro3     \\
Mrk 178        & 11\ \ 33\ \ 28.9~        &$+49$\ 14\ 13                      &15.0   &0.00076&--12.4&UGC 4561  \\
SBS $1152+579$ & 11\ \ 55\ \ 28.3~        &$+57$\ 39\ 52                      &16.0   &0.01732&--18.2&Mrk 193   \\
Mrk 209        & 12\ \ 26\ \ 16.1~        &$+48$\ 29\ 31                      &15.0   &0.00101&--13.0&I Zw 36   \\
Mrk 59         & 12\ \ 59\ \ 00.3~        &$+34$\ 50\ 40                      &13.0   &0.00258&--17.1&NGC 4861  \\
J $1404+5423$  & 14\ \ 04\ \ 29.5~        &$+54$\ 23\ 47                      &14.0   &0.00095&--13.9&          \\
\multicolumn{7}{c}{b) Galaxies observed with the 4m KPNO telescope
with detected [Fe {\sc v}] $\lambda$4227 emission} \\
J $0519+0007$  & 05\ \ 19\ \ 02.7~        &$+00$\ 07\ 29                      &18.2   &0.04460&--18.1&          \\
HS $1028+3843$ & 10\ \ 31\ \ 51.8~        &$+38$\ 28\ 07                      &19.4   &0.02945&--16.0&          \\
HS $2236+1344$ & 22\ \ 38\ \ 31.1~        &$+14$\ 00\ 29                      &18.2   &0.02115&--16.4&          \\
\multicolumn{7}{c}{c) SDSS galaxies with detected 
[Fe {\sc v}] $\lambda$ 4227 emission} \\
J $0240-0828$ & 02\ \ 40\ \ 52.2~         &$-08$\ 28\ 27                      &19.8   &0.08231&--18.1&          \\
J $0840+4707$ & 08\ \ 40\ \ 29.9~         &$+47$\ 07\ 09                      &18.0   &0.04212&--18.1&HS $0837+4717$ \\
J $0944-0038$  & 09\ \ 44\ \ 01.9~        &$-00$\ 38\ 32                      &17.5   &0.00483&--13.9&CGCG 007$-$025 \\
J $1253-0312$  & 12\ \ 53\ \ 06.0~        &$-03$\ 12\ 59                      &16.2   &0.02280&--18.6&          \\
J $1323-0132$  & 13\ \ 23\ \ 47.5~        &$-01$\ 32\ 52                      &19.1   &0.02254&--15.7&          \\
\enddata 
\end{deluxetable}

%
\begin{deluxetable}{lccc}
\small
\tablenum{2}
\tablecolumns{4}
\tablewidth{0pt}
\tablecaption{Journal of the MMT Observations}
\tablehead{
\colhead{Name}&\colhead{Date of Obs.}&\colhead{Exposure}&\colhead{Airmass}
}
\startdata
SBS $0335-052$W & 2005, Feb 4       & 2700s & 1.27 - 1.31   \\
SBS $0335-052$E & 2004, Feb 19 - 20 & 6300s & 1.32 - 1.47   \\
J $0519+0007$   & 2004, Feb 19      & 3600s & 1.20 - 1.28   \\
II Zw 40        & 2005, Feb 4       & 1200s & 1.32 - 1.37   \\
Mrk 71          & 2004, Feb 19 - 20 & 1320s & 1.26 - 1.37   \\
HS $0822+3542$  & 2005, Feb 4       & 2700s & 1.12 - 1.19   \\
Mrk 94          & 2005, Feb 4       & 1800s & 1.32 - 1.36   \\
HS $0837+4717$  & 2004, Feb 20      & 2700s & 1.13 - 1.19   \\
SBS $0911+472$  & 2005, Feb 4       & 1800s & 1.04 - 1.05   \\
SBS $0926+606$A & 2005, Feb 4       & 1800s & 1.18 - 1.20   \\
I Zw 18         & 2005, Feb 4       & 3600s & 1.19 - 1.28   \\
SBS $0940+544$  & 2005, Feb 4       & 2700s & 1.08 - 1.09   \\
SBS $1030+583$  & 2005, Feb 4       & 1800s & 1.12 - 1.13   \\
Mrk 35          & 2004, Feb 20      & 1575s & 1.10 - 1.12   \\
Mrk 178         & 2005, Feb 4       & 1800s & 1.06 - 1.07   \\
SBS $1152+579$  & 2005, Feb 4       & 1178s & 1.12 - 1.13   \\
Mrk 209         & 2004, Feb 19      & 2400s & 1.14 - 1.20   \\
Mrk 59          & 2004, Feb 20      & 1200s & 1.06 - 1.09   \\
J $1404+5423$   & 2005, Feb 4       &  600s & 1.09          \\
\enddata 
\end{deluxetable}

\clearpage


\begin{figure}
\epsscale{1.1}
\plotone{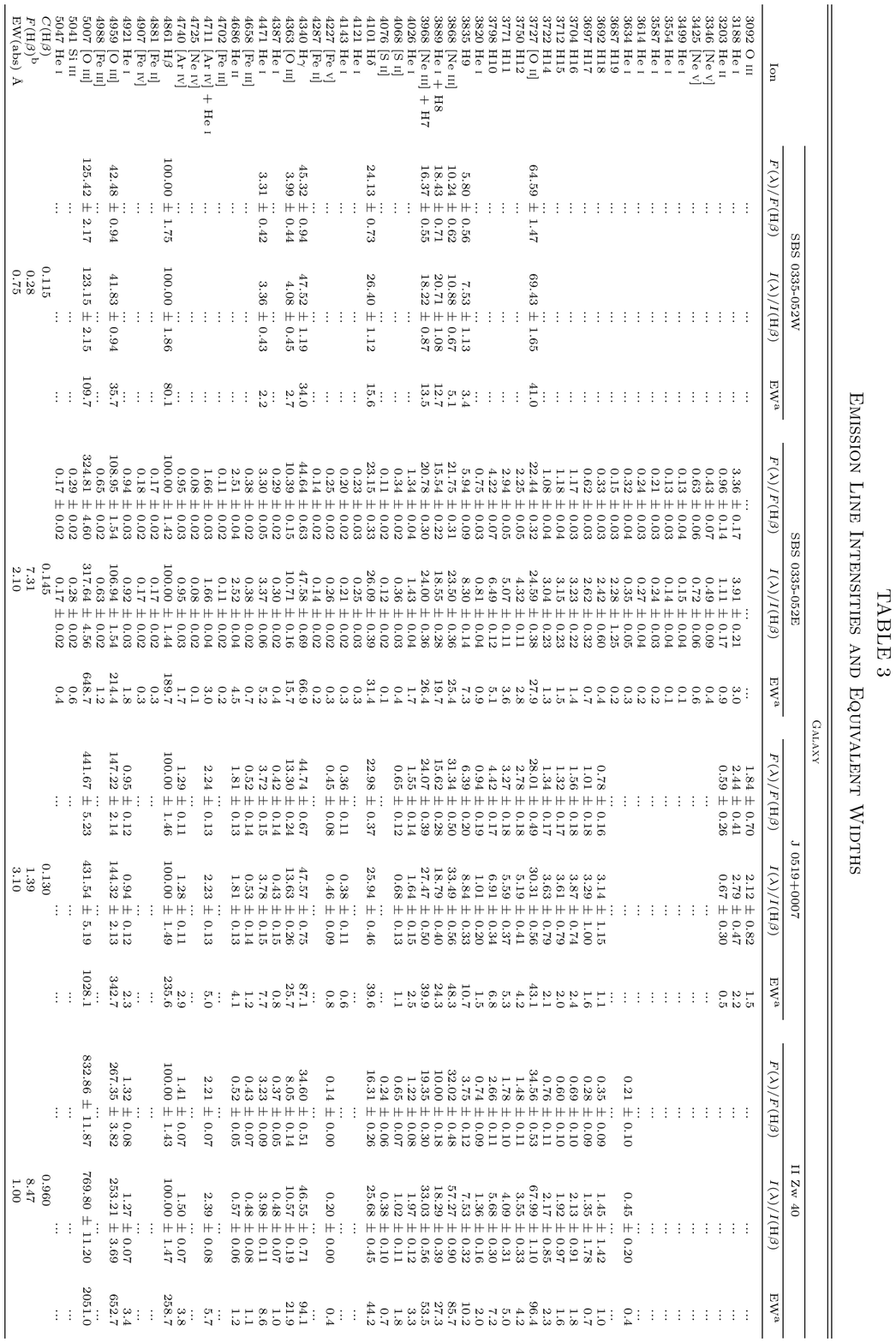}
\end{figure}

\clearpage


\begin{figure}
\epsscale{1.1}
\plotone{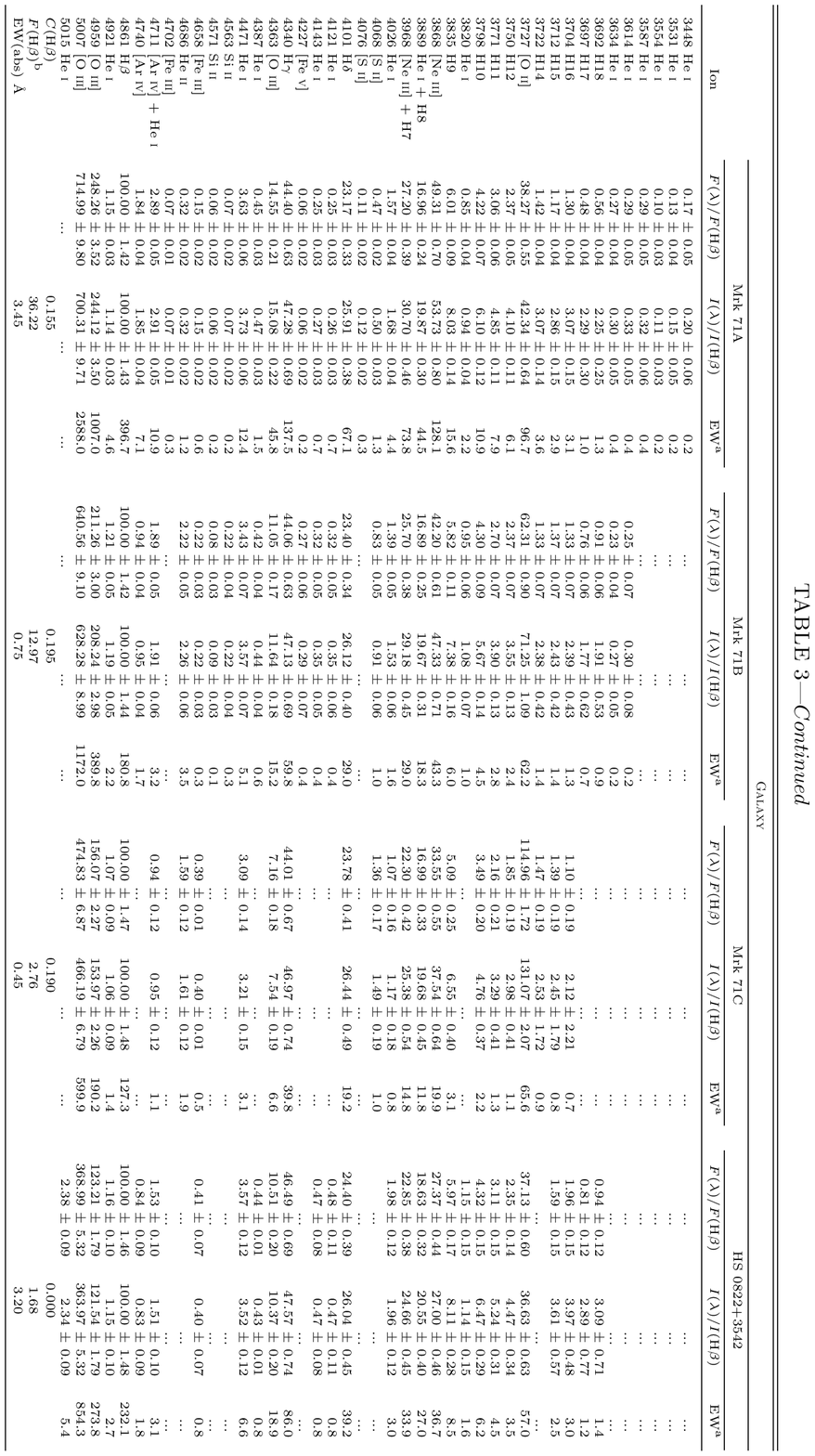}
\end{figure}

\clearpage


\begin{figure}
\epsscale{1.1}
\plotone{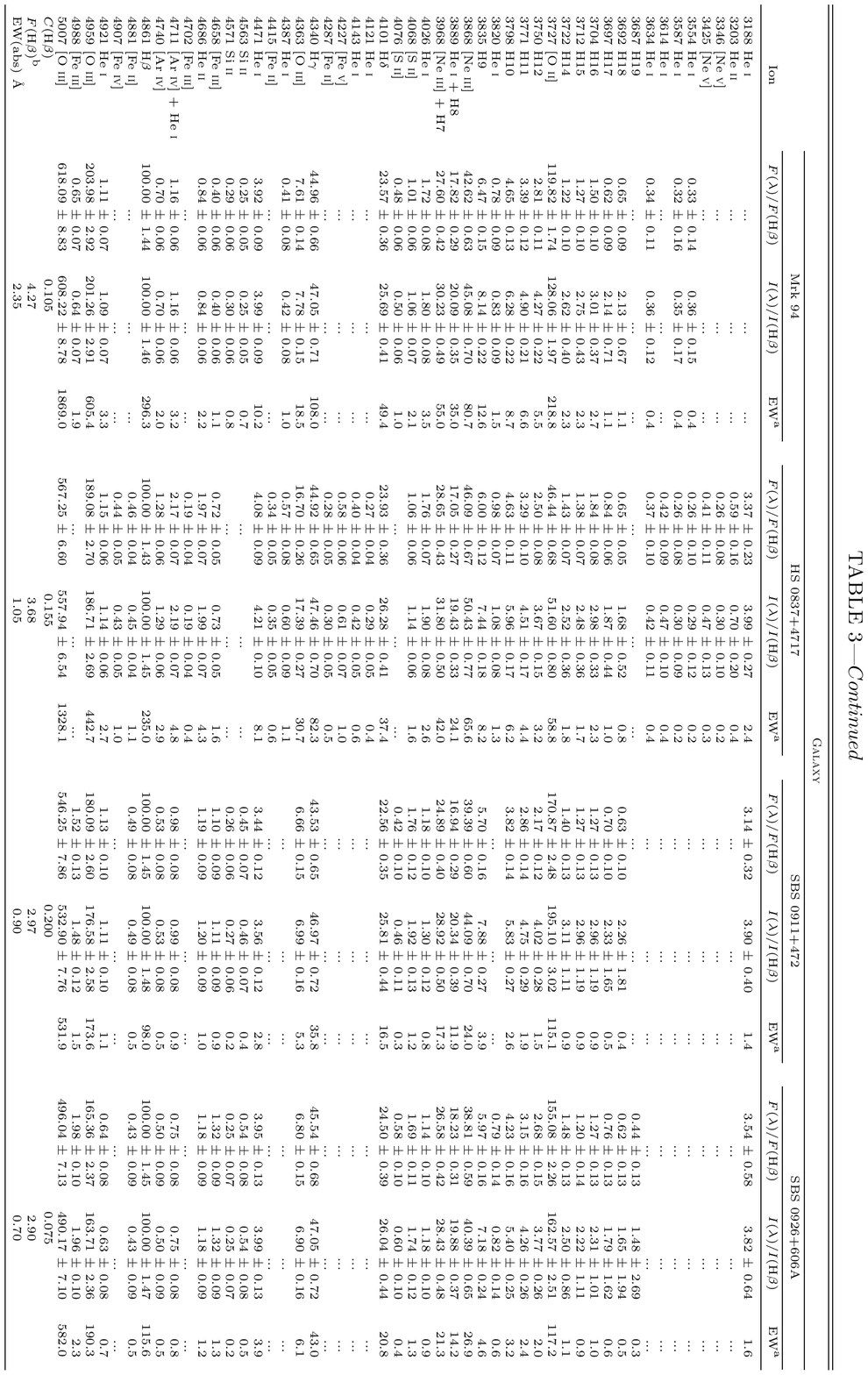}
\end{figure}

\clearpage


\begin{figure}
\epsscale{1.1}
\plotone{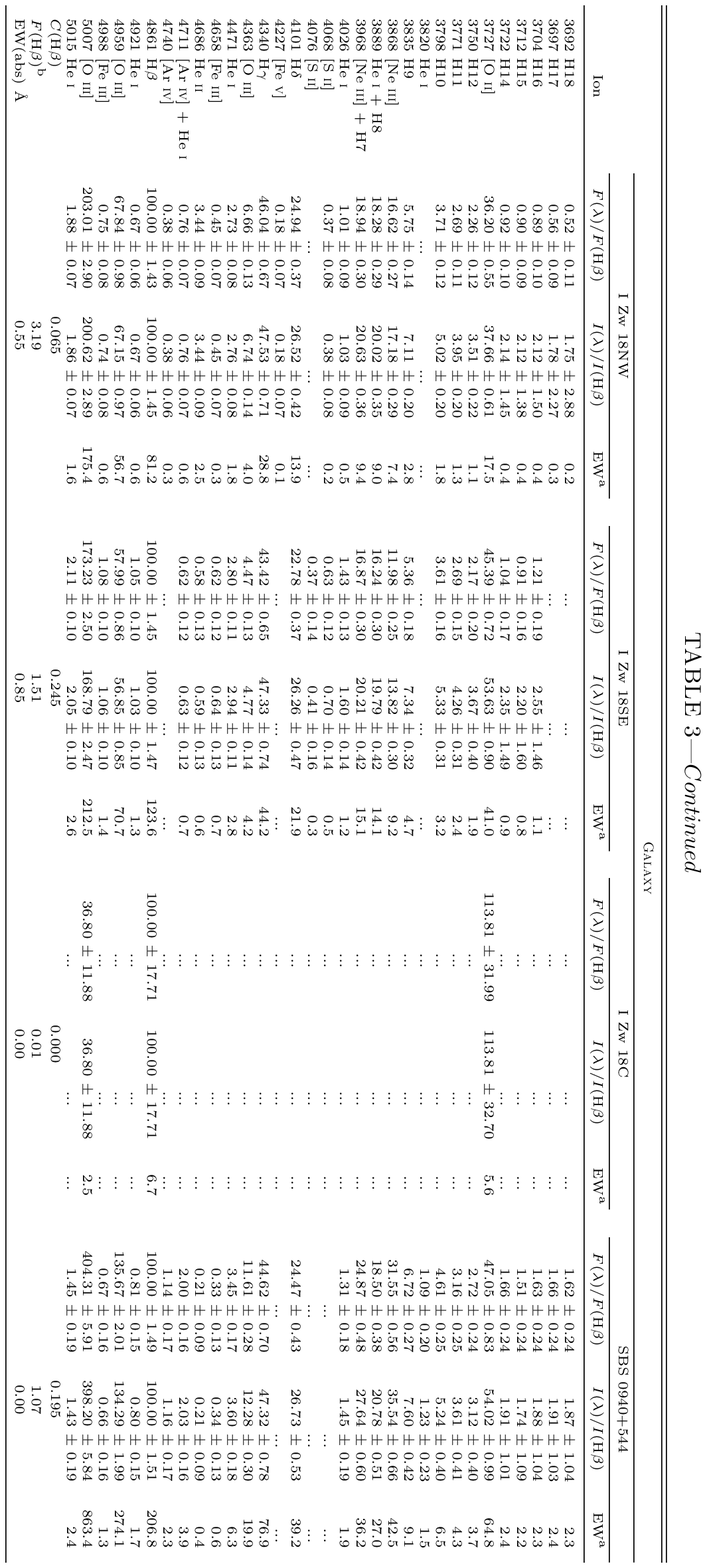}
\end{figure}

\clearpage


\begin{figure}
\epsscale{1.1}
\plotone{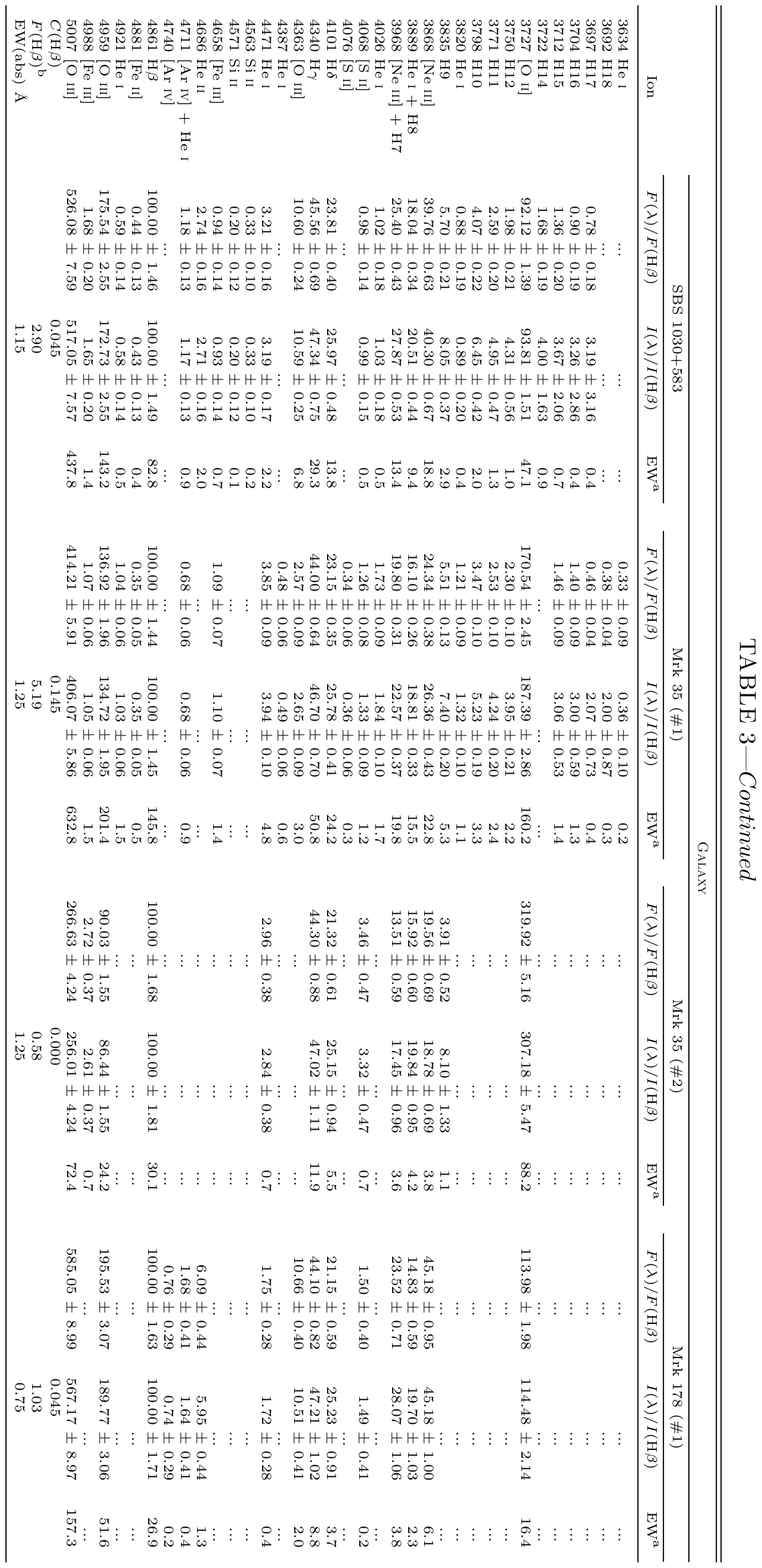}
\end{figure}

\clearpage


\begin{figure}
\epsscale{1.1}
\plotone{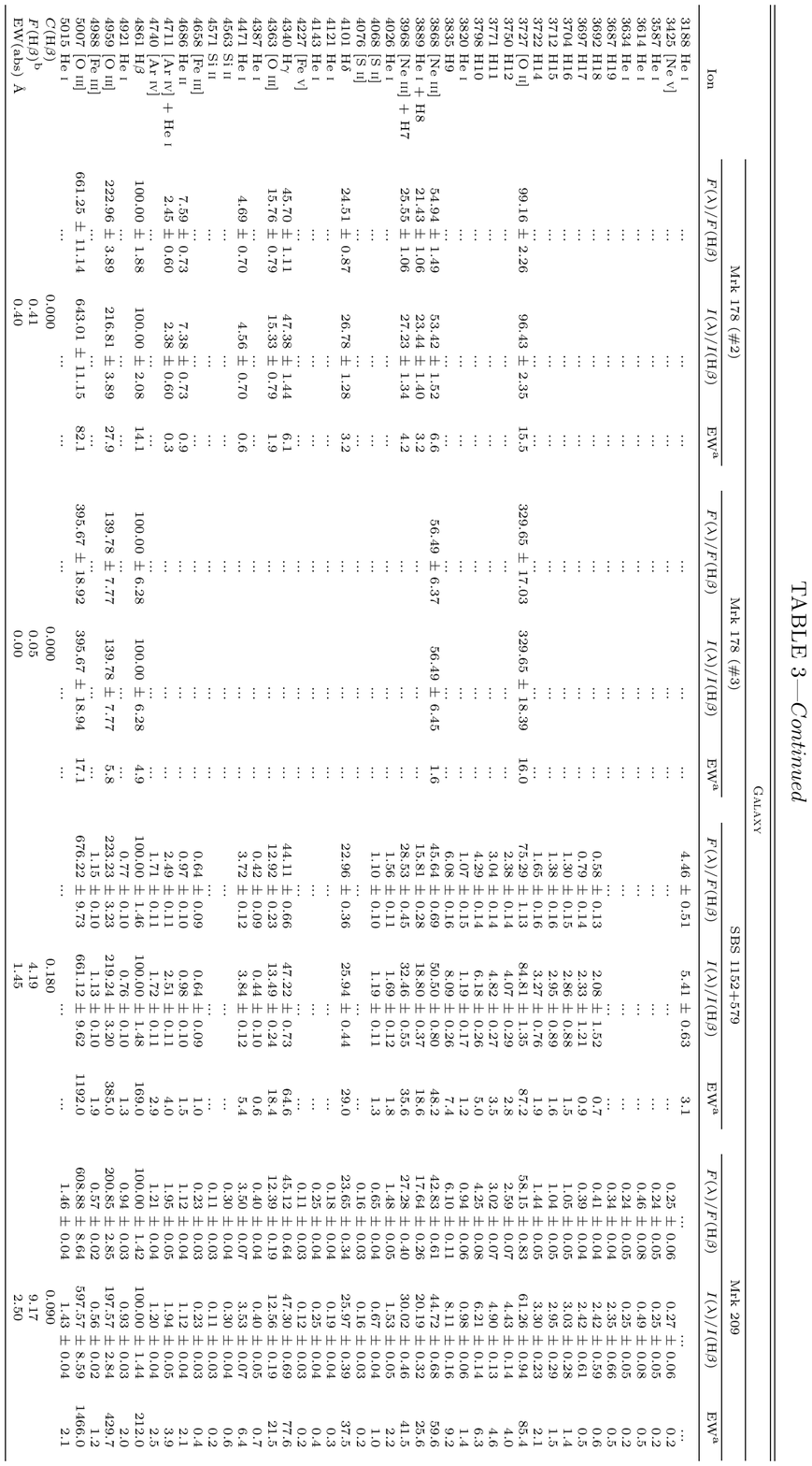}
\end{figure}

\clearpage


\begin{figure}
\epsscale{1.1}
\plotone{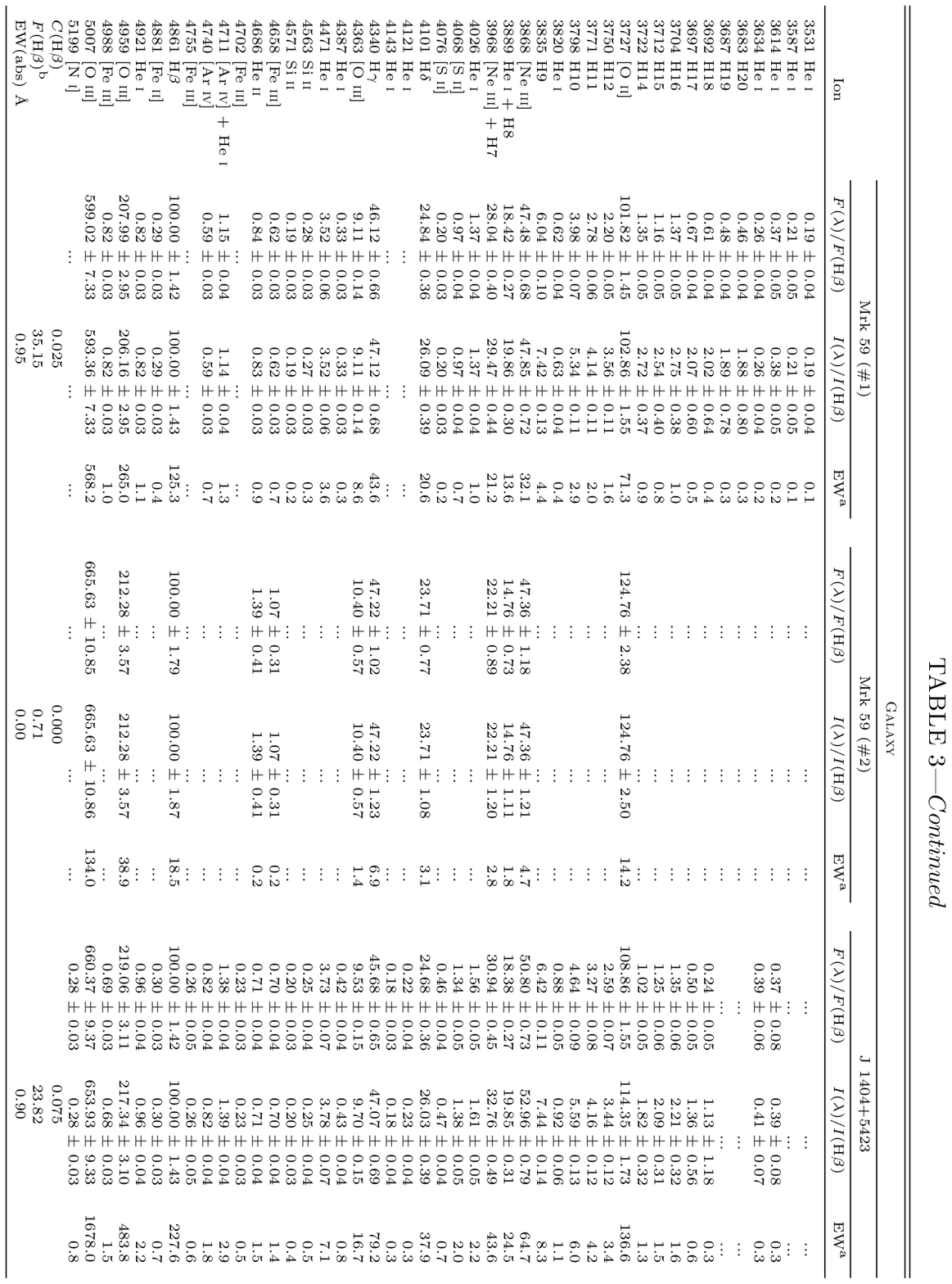}
\end{figure}

\clearpage


\begin{figure}
\epsscale{1.1}
\plotone{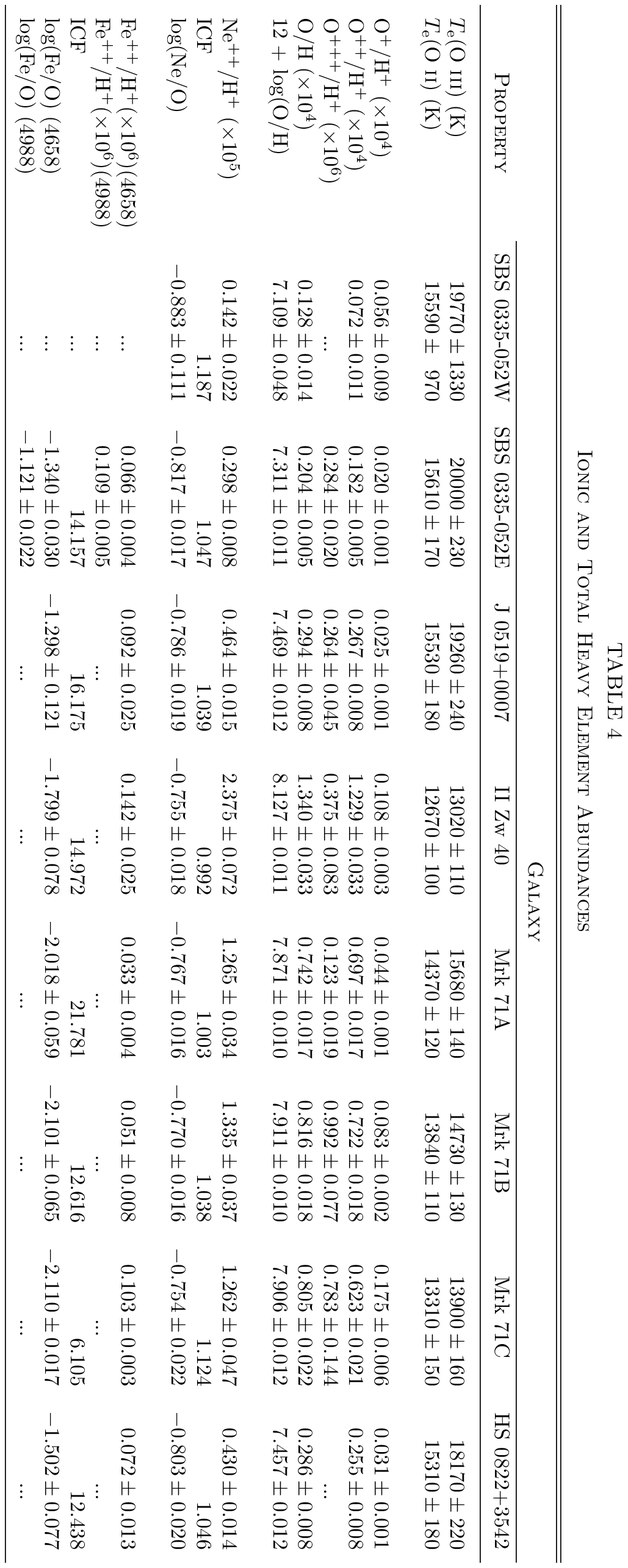}
\end{figure}

\clearpage


\begin{figure}
\epsscale{1.1}
\plotone{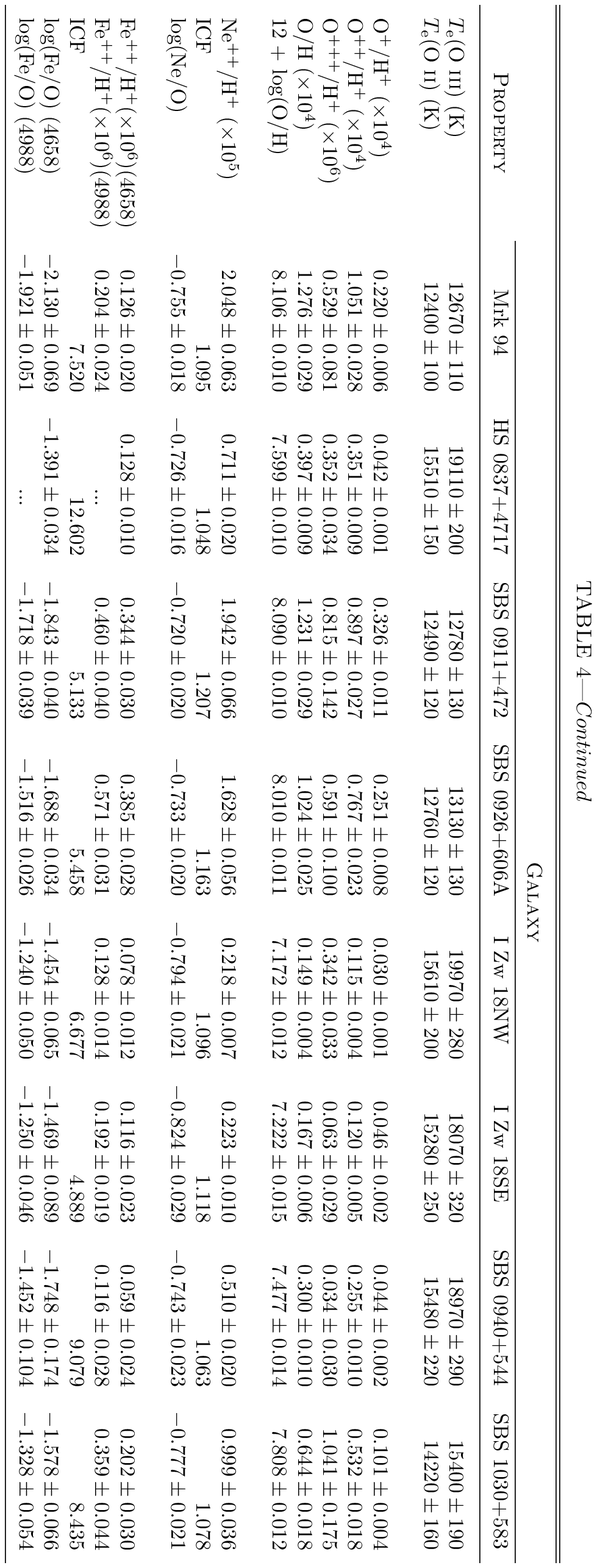}
\end{figure}

\clearpage


\begin{figure}
\epsscale{1.1}
\plotone{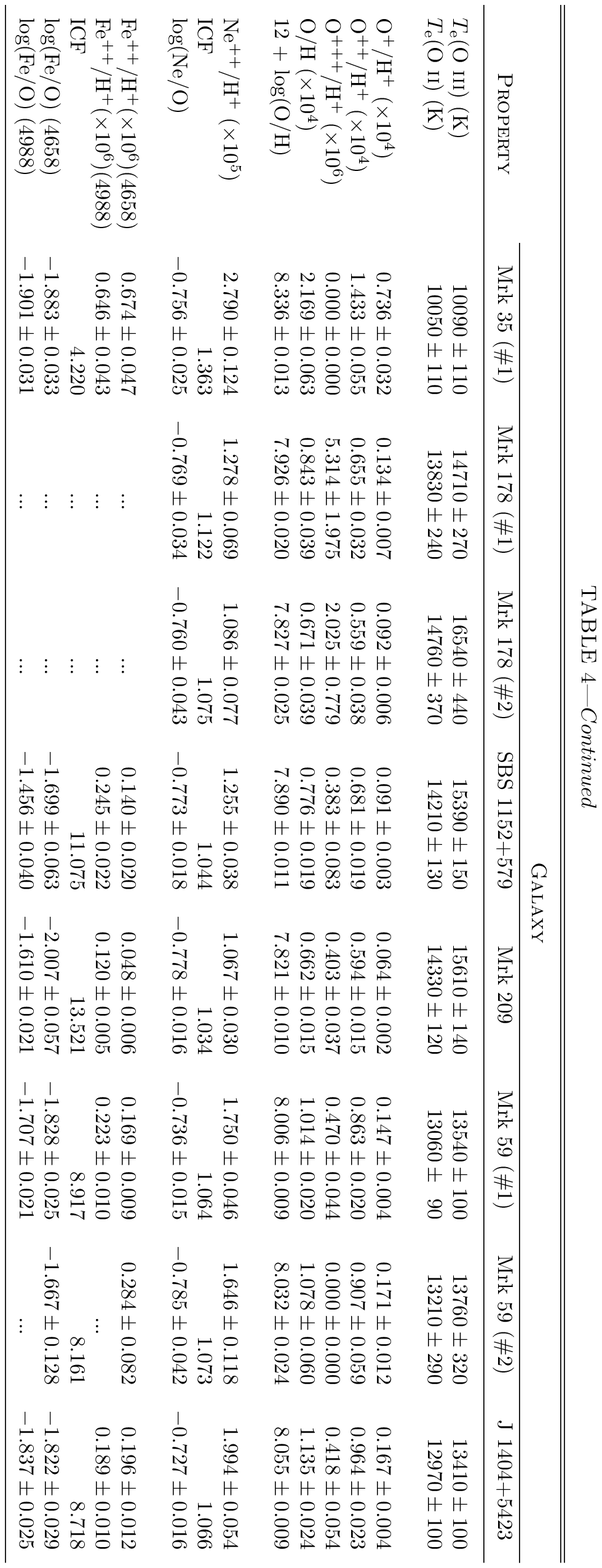}
\end{figure}

\clearpage


\begin{figure}
\epsscale{1.1}
\plotone{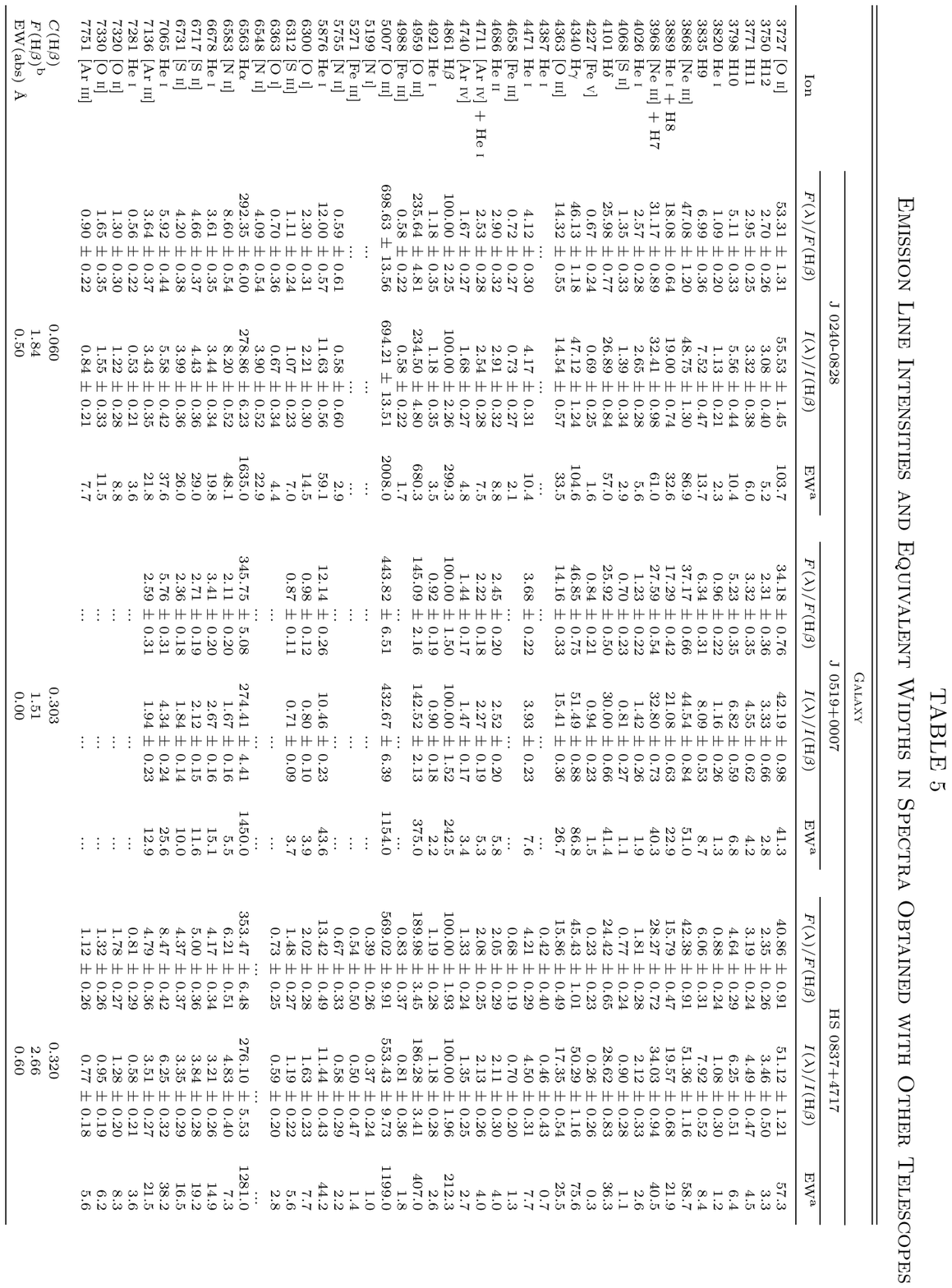}
\end{figure}

\clearpage


\begin{figure}
\epsscale{1.1}
\plotone{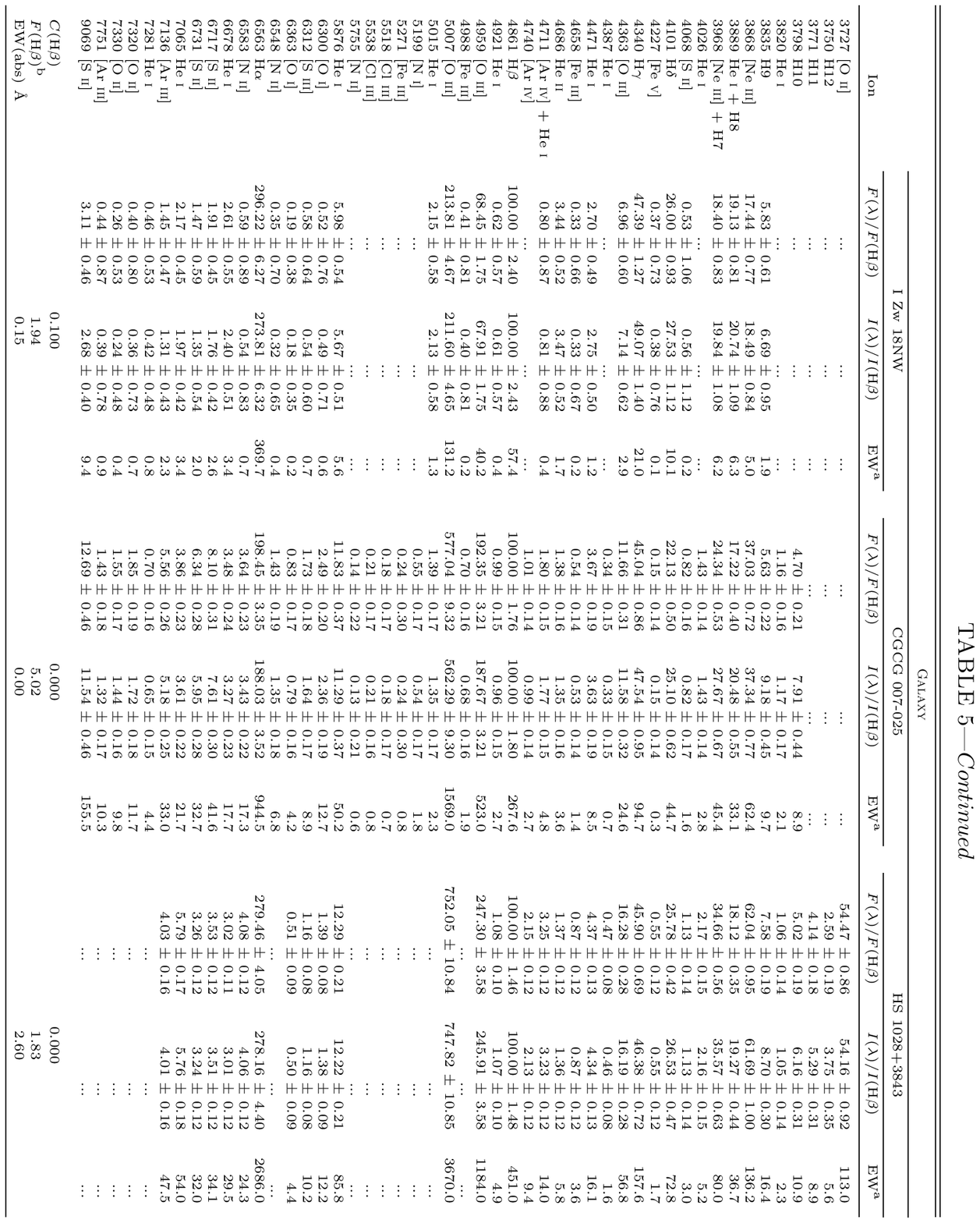}
\end{figure}

\clearpage


\begin{figure}
\epsscale{1.1}
\plotone{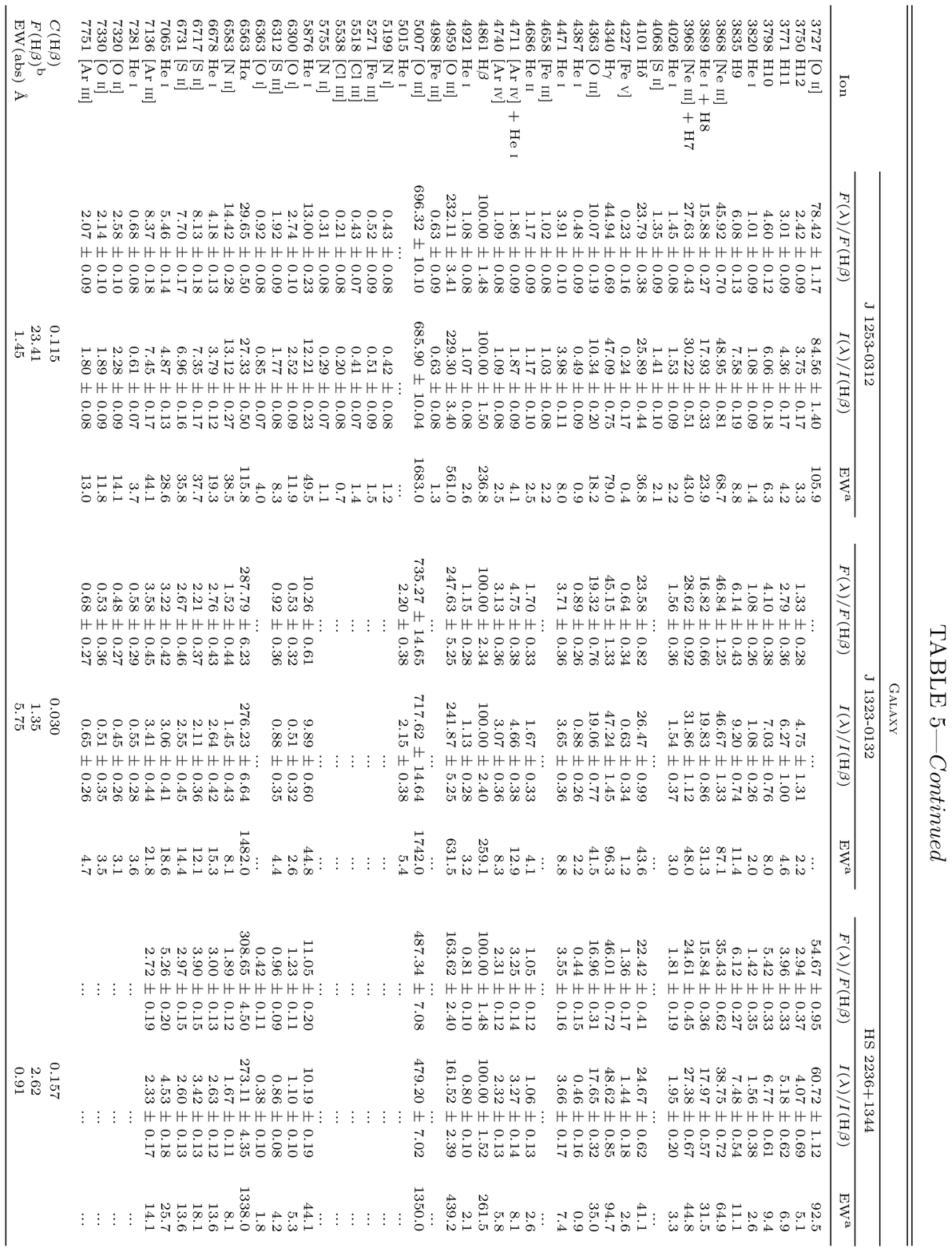}
\end{figure}

\clearpage


\begin{figure}
\epsscale{1.1}
\plotone{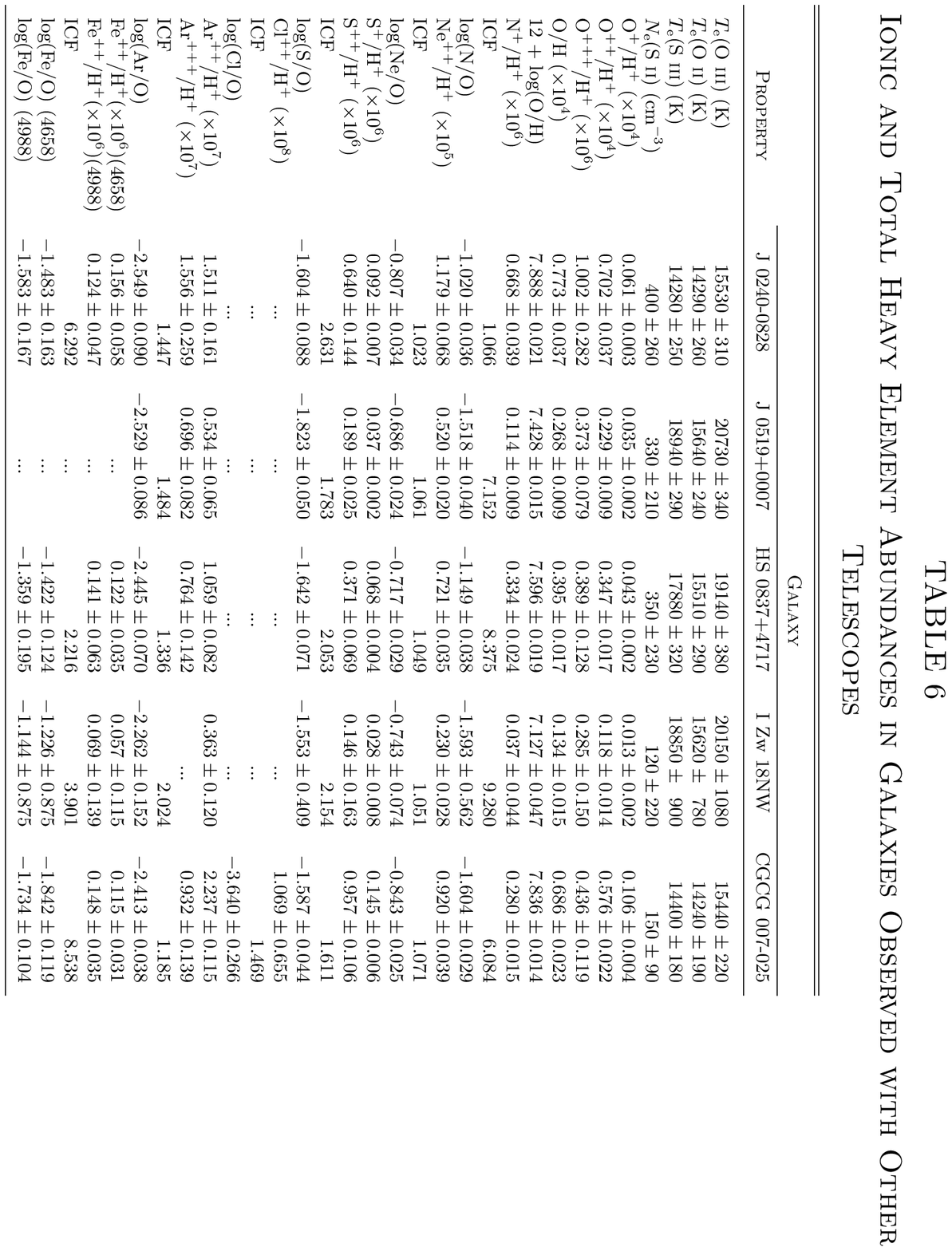}
\end{figure}

\clearpage


\begin{figure}
\epsscale{1.1}
\plotone{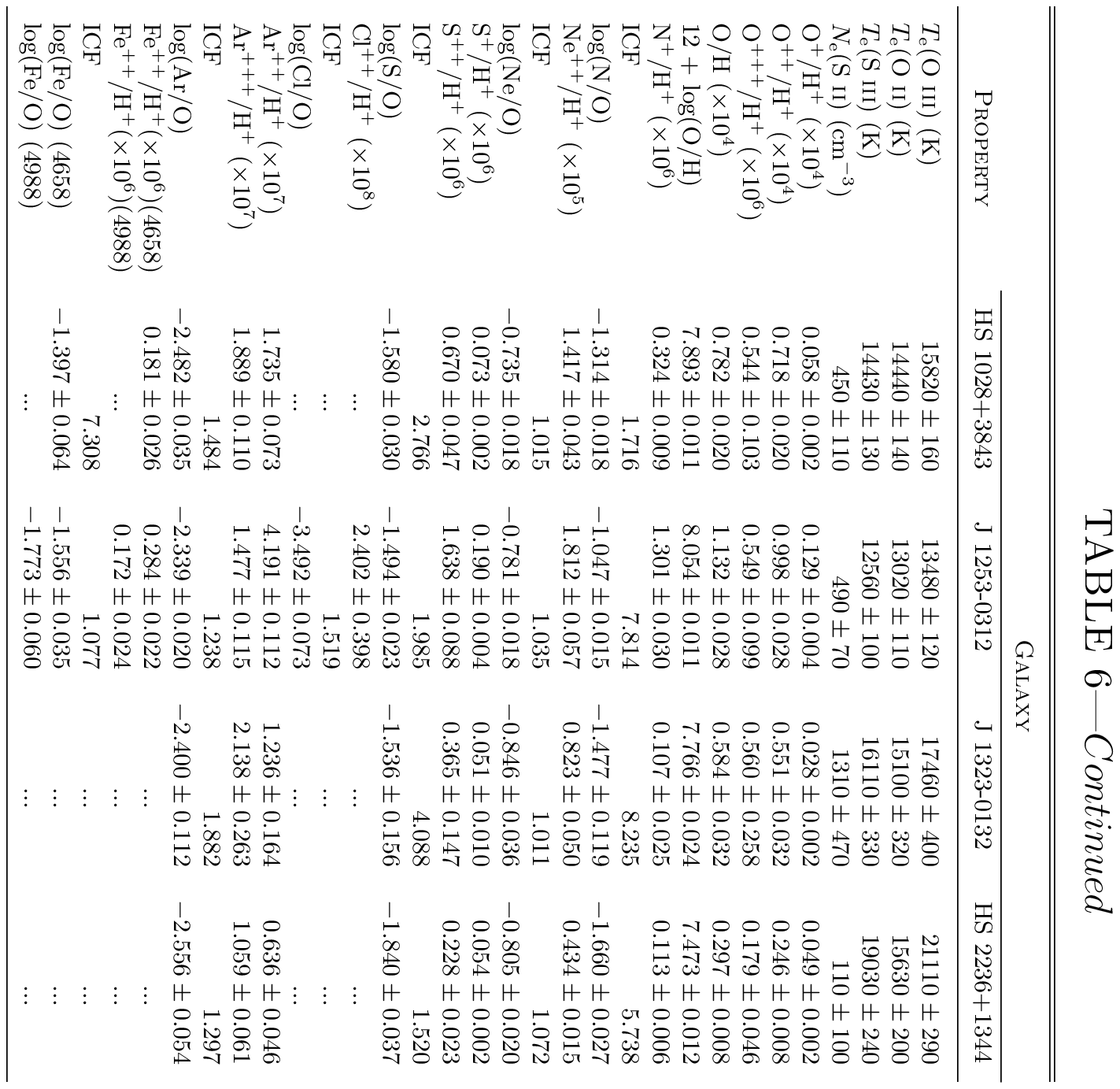}
\end{figure}

%
\begin{deluxetable}{lccccc}
\tablenum{7}
\tablecolumns{6}
\tablewidth{0pt}
\tablecaption{Parameters of high ionization lines}
\tablehead{
\colhead{Name}&\colhead{12 + log O/H}
&\colhead{$\frac{I({\rm He\ II}\ 4686)}{I({\rm He\ I}\ 4471)}$}
&\colhead{$\frac{I([{\rm Ne\ V}]\ 3426)}{I([{\rm Ne\ III}]\ 3868)}$}
&\colhead{$\frac{I([{\rm Fe\ V}]\ 4227)}{I([{\rm Fe\ III}]\ 4658)}$}
&\colhead{$\frac{I([{\rm Fe\ V}]\ 4227)}{I([{\rm Fe\ III}]\ 4988)}$}
}
\startdata
I Zw 18NW (SDSS)       & 7.13 & 1.42 &\nodata& 1.15  & 0.95  \\
I Zw 18NW (MMT)        & 7.17 & 1.24 &\nodata& 0.40  & 0.24  \\
SBS $0335-052$E (MMT)  & 7.31 & 0.75 & 0.031 & 0.68  & 0.41  \\
J $0519+0007$ (4m)     & 7.43 & 0.64 &\nodata&\nodata&\nodata\\
J $0519+0007$ (MMT)    & 7.47 & 0.48 &\nodata& 0.87  &\nodata\\
HS $2238+1344$ (4m)    & 7.47 & 0.29 &\nodata&\nodata&\nodata\\
Tol $1214-277$ (VLT)   & 7.55 & 1.46 &\nodata& 1.74  & 1.48  \\
Tol $1214-277$ (3.6m)  & 7.56 & 1.67 & 0.085 &\nodata&\nodata\\
HS $0837+4717$ (SDSS)  & 7.60 & 0.47 &\nodata& 0.37  & 0.32  \\
HS $0837+4717$ (MMT)   & 7.60 & 0.47 & 0.009 & 0.84  &\nodata\\
J $1323-0132$ (SDSS)   & 7.77 & 0.46 &\nodata&\nodata&\nodata\\
Mrk 209 (MMT)          & 7.82 & 0.32 & 0.006 & 0.52  & 0.21  \\
CGCG $007-025$ (SDSS)  & 7.84 & 0.37 &\nodata& 0.28  & 0.22  \\
Mrk 71A (MMT)          & 7.87 & 0.09 &\nodata& 0.40  &\nodata\\
HS $1028+3843$ (4m)    & 7.89 & 0.31 &\nodata& 0.63  &\nodata\\
J $0240-0828$ (SDSS)   & 7.89 & 0.70 &\nodata& 0.95  & 1.19  \\
Mrk 71B (MMT)          & 7.91 & 0.64 &\nodata& 1.32  &\nodata\\
J $1253-0312$ (SDSS)   & 8.05 & 0.29 &\nodata& 0.23  & 0.38  \\
II Zw 40 (MMT)         & 8.12 & 0.14 &\nodata& 0.42  &\nodata\\ \hline
mean [12+log(O/H) $\leq$ 7.6]&& 0.89$\pm$0.49 &       & 0.86$\pm$0.44  & 0.68$\pm$0.47  \\
mean [12+log(O/H) $>$ 7.6]   && 0.37$\pm$0.19 &       & 0.59$\pm$0.35  & 0.50$\pm$0.40  \\
\enddata 
\end{deluxetable}

\clearpage


\begin{figure}
\figurenum{1}
\epsscale{0.9}
\plotone{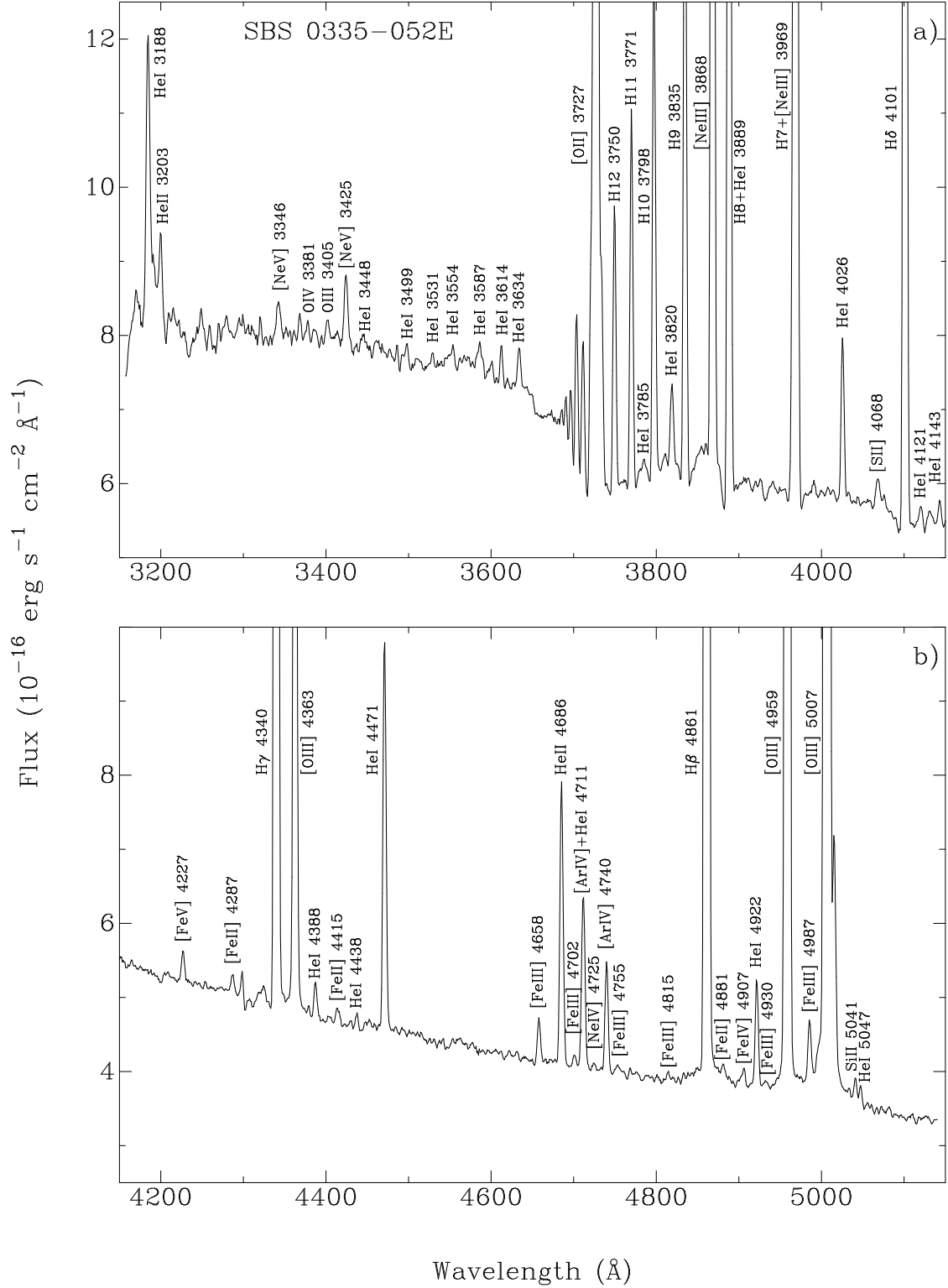}
\caption{MMT spectrum of SBS 0335--052E with labeled emission lines and
corrected to the rest frame.
\label{Fig1}}
\end{figure}

\clearpage


\begin{figure}
\figurenum{2}
\epsscale{0.9}
\plotone{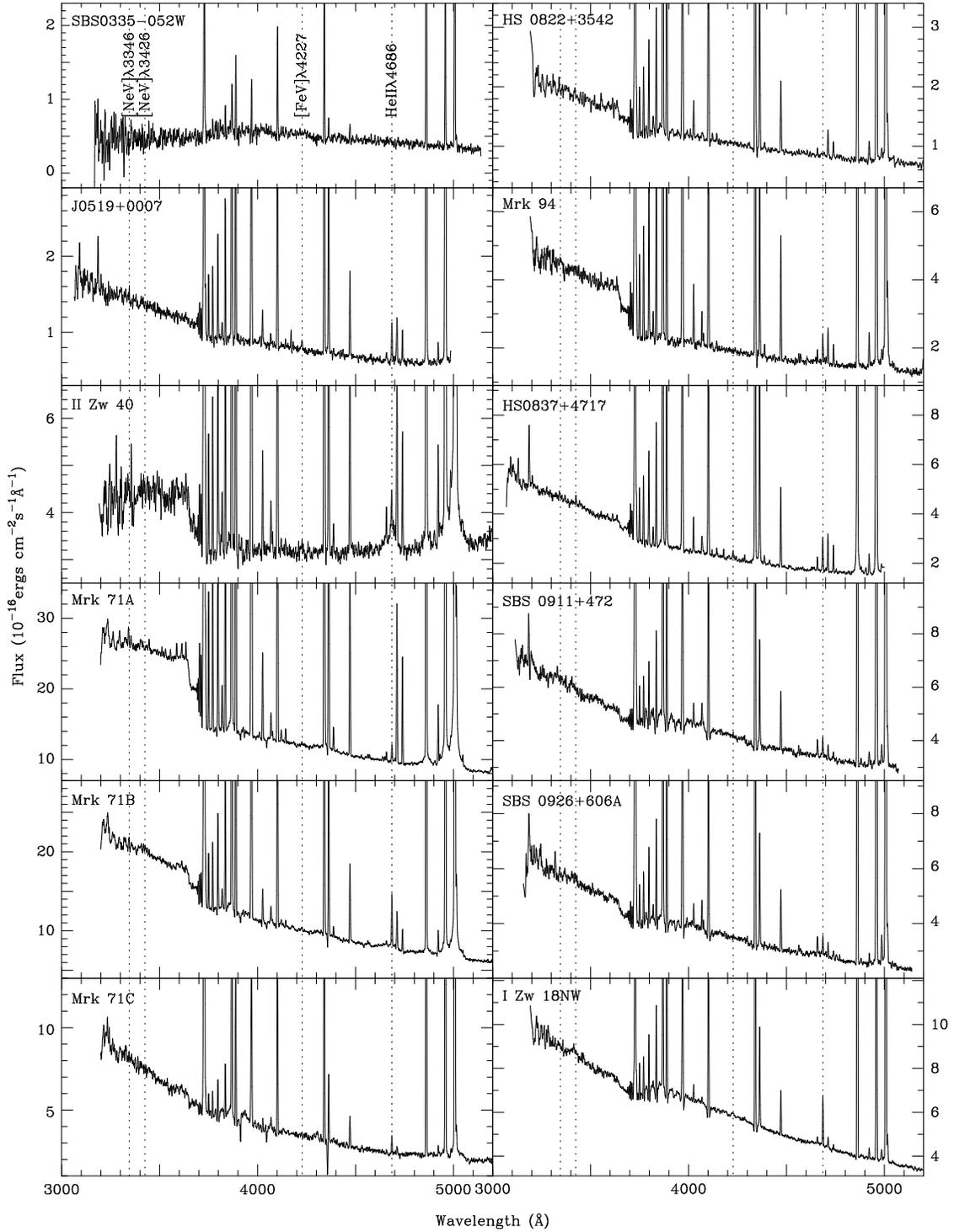}
\caption{Spectra of other galaxies observed with the MMT corrected to the
rest frame.
The locations of the high-ionization [Ne {\sc v}] $\lambda$3346, 3426,
[Fe {\sc v}] $\lambda$4227 and He {\sc ii} $\lambda$4686 emission lines are 
indicated by vertical dotted lines. The spectra are given from top to
bottom and from left to right in order of increasing right ascension,
as in Table \ref{Tab1}.
\label{Fig2}}
\end{figure}

\clearpage


\begin{figure}
\figurenum{2}
\epsscale{0.9}
\plotone{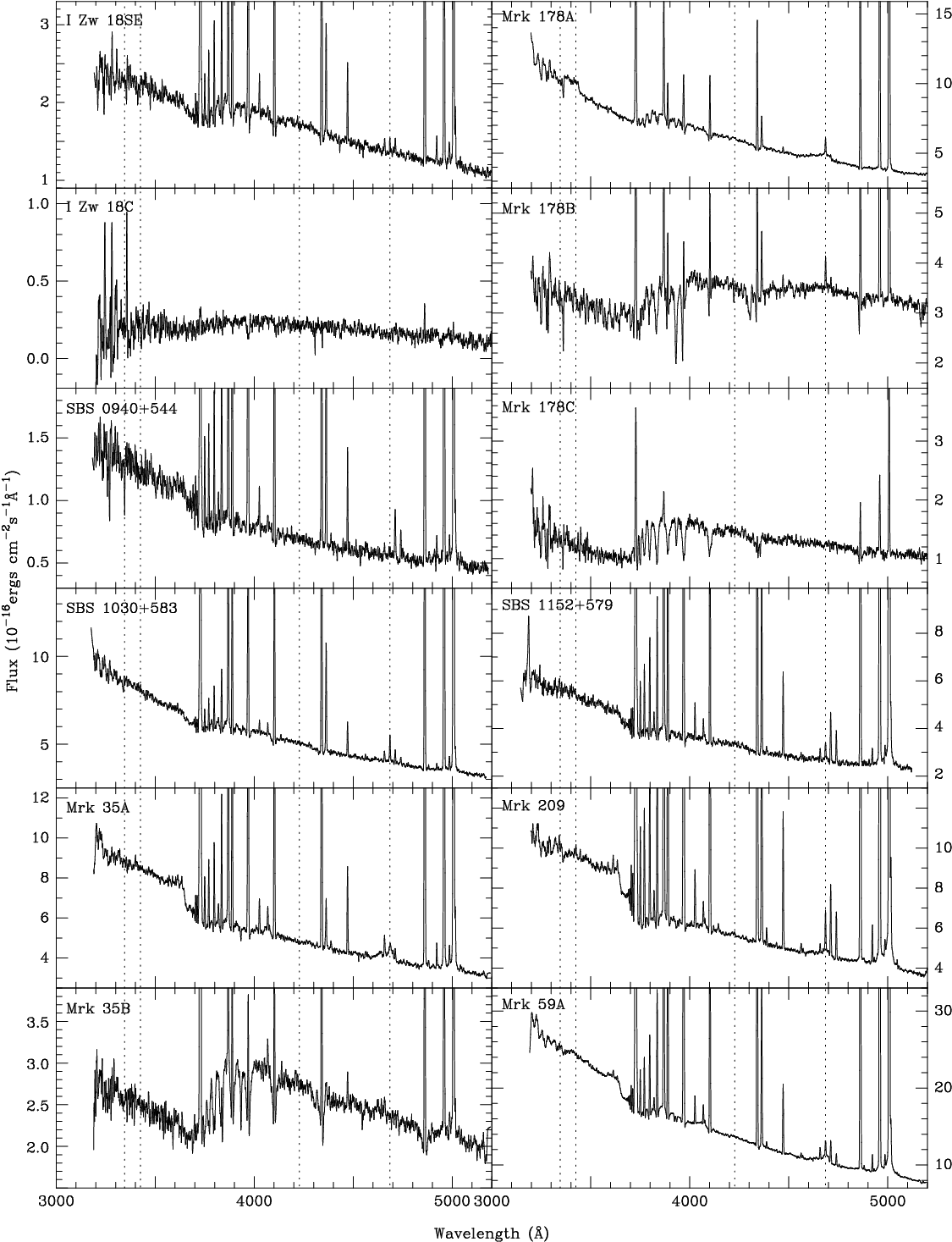}
\caption{Continued.
}
\end{figure}

\clearpage


\begin{figure}
\figurenum{2}
\epsscale{0.9}
\plotone{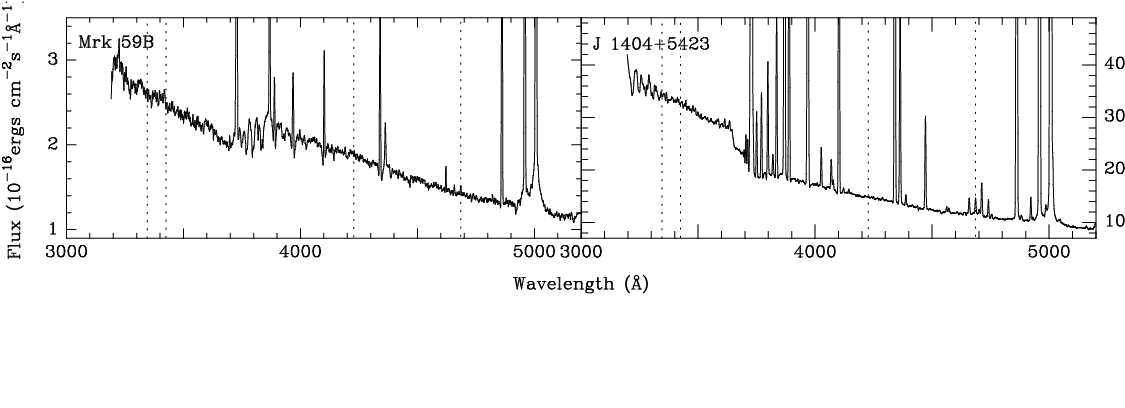}
\caption{Continued.
}
\end{figure}

\clearpage


\begin{figure}
\figurenum{3}
\epsscale{0.9}
\plotone{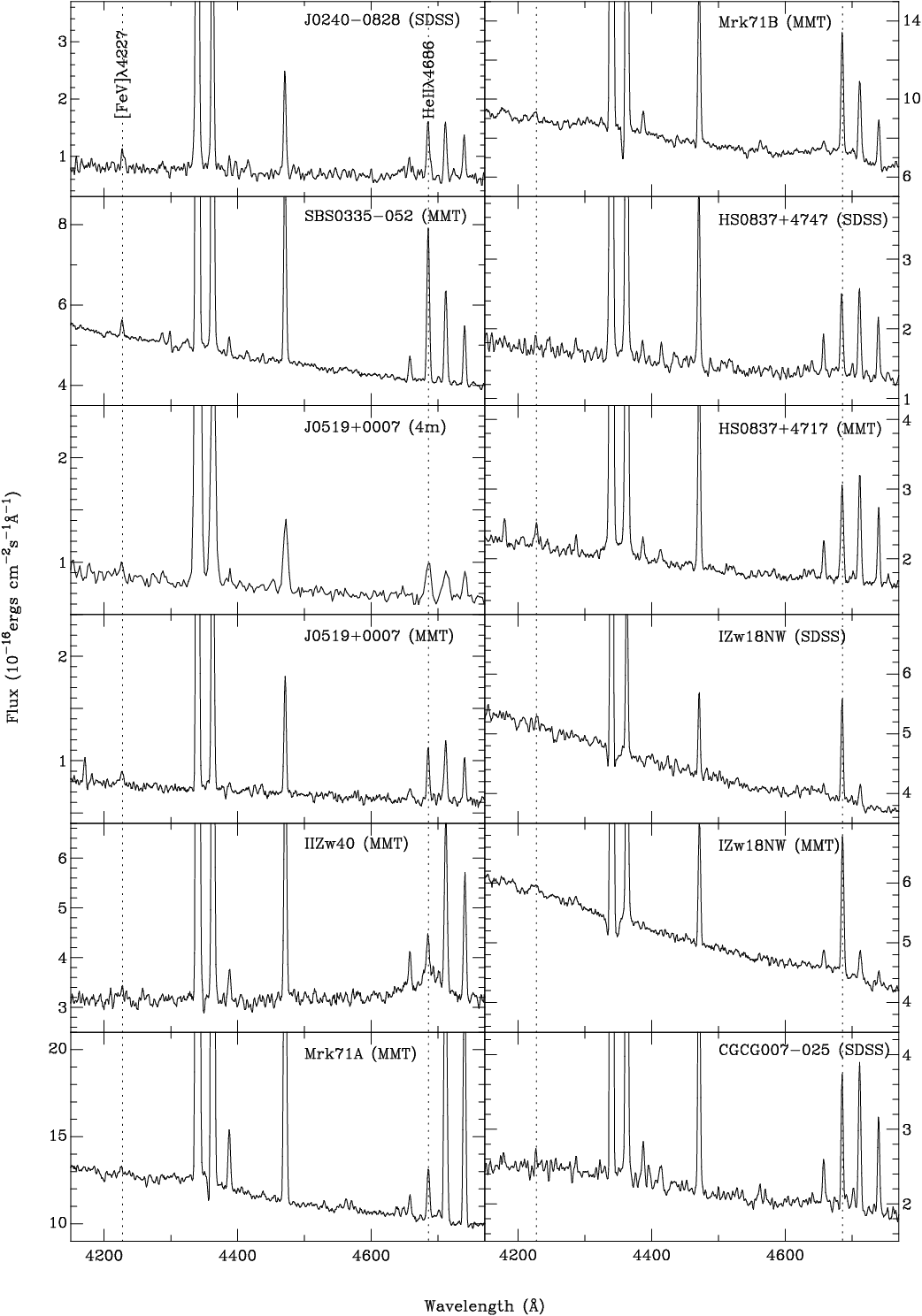}
\caption{Spectra of galaxies with a detected [Fe {\sc v}] $\lambda$4227
emission line. The vertical dotted lines show the location of the 
[Fe {\sc v}] $\lambda$4227 and He {\sc ii} $\lambda$4686 emission lines.
\label{Fig3}}
\end{figure}

\clearpage


\begin{figure}
\figurenum{3}
\epsscale{0.9}
\plotone{fig3_2.ps}
\caption{Continued.
}
\end{figure}

\clearpage


\begin{figure}
\figurenum{4}
\epsscale{0.9}
\plotone{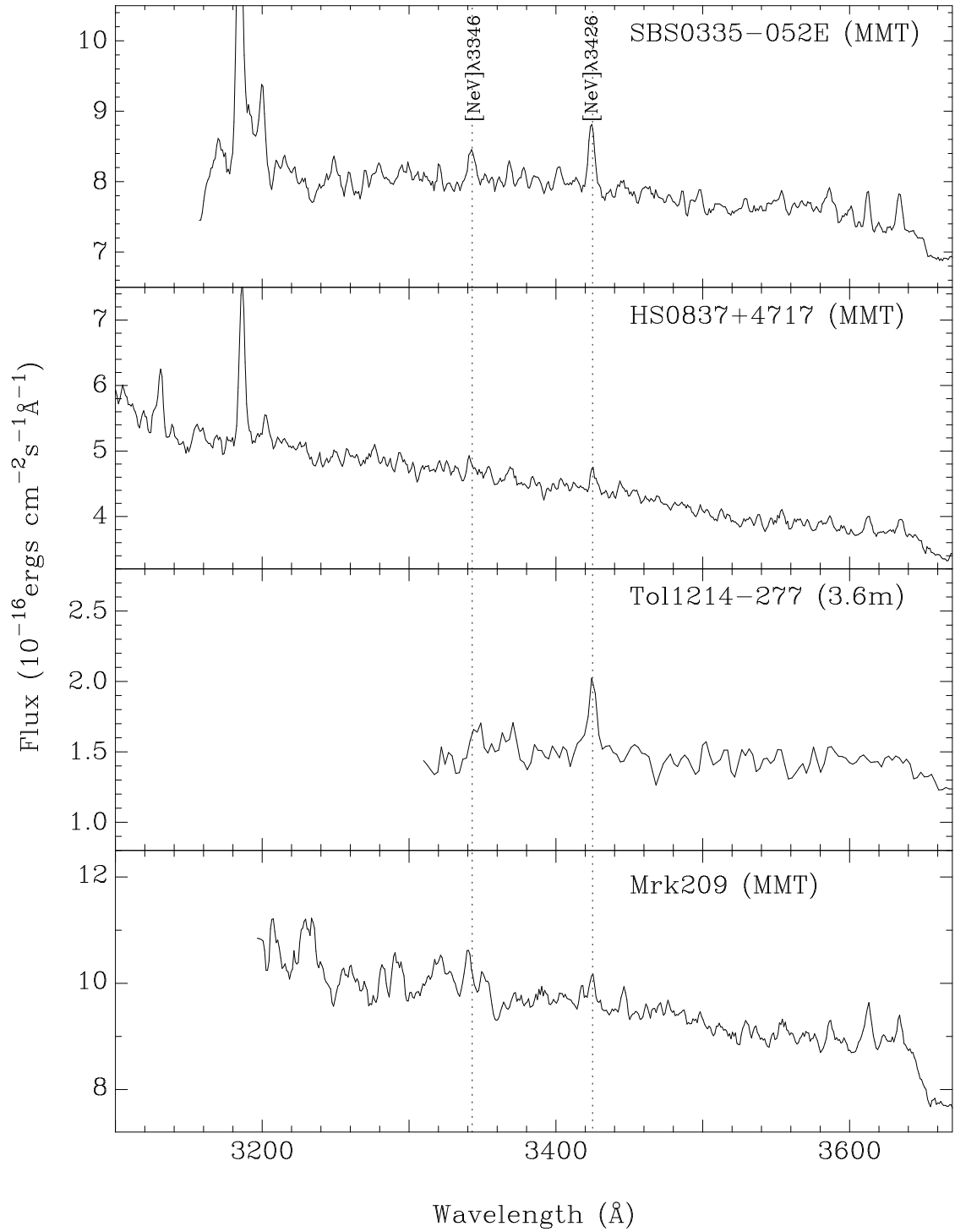}
\caption{Spectra of galaxies with a detected [Ne {\sc v}] $\lambda$3426
emission line. The locations of this line and of the
[Ne {\sc v}] $\lambda$3346 emission line are indicated by dotted vertical 
lines.
\label{Fig4}}
\end{figure}

\clearpage

\begin{figure}
\figurenum{5}
\plotfiddle{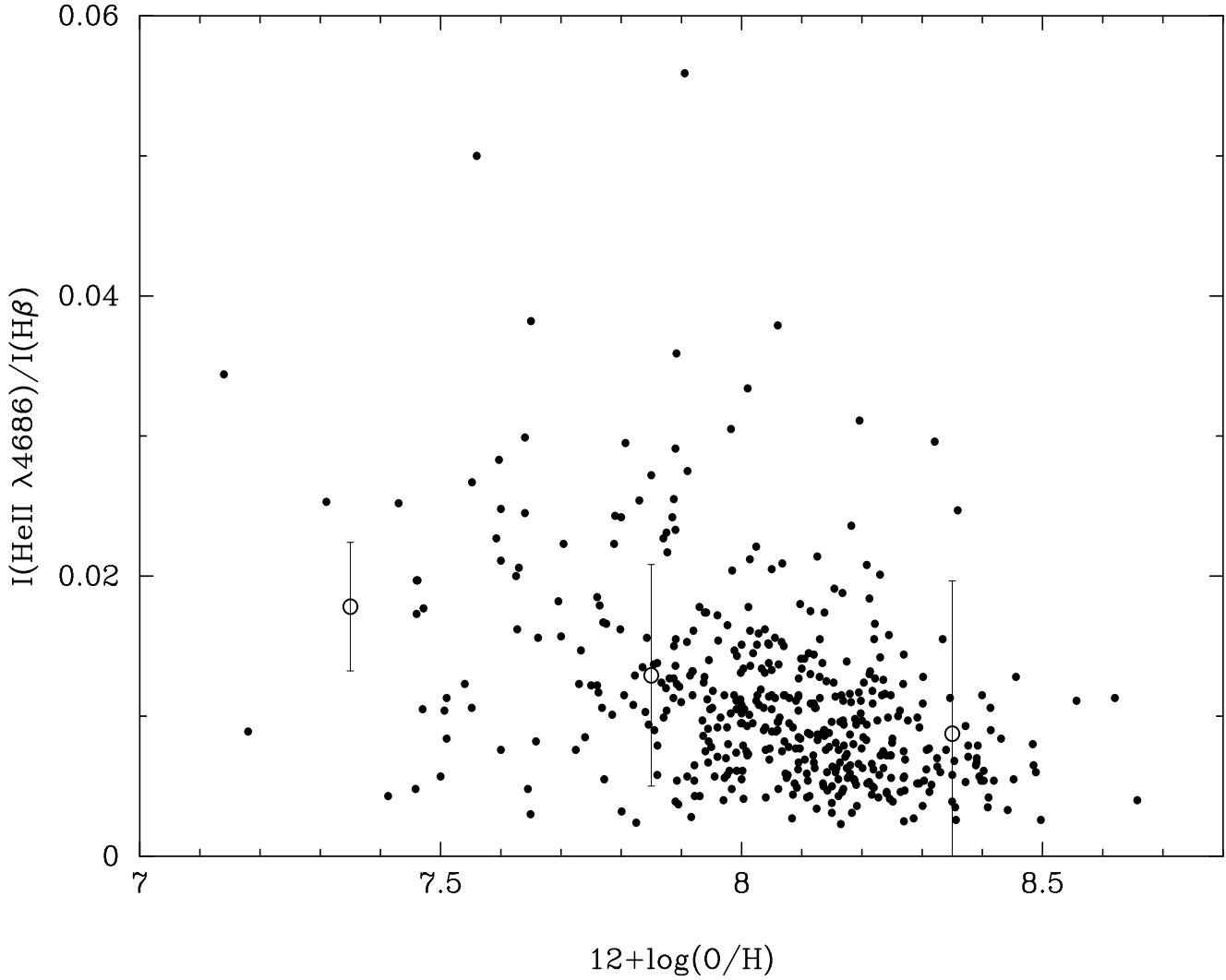}{1pt}{-90.}{400.}{500.}{-40.}{0.}
\caption{Intensity of the He {\sc ii} $\lambda$4686
emission line relative to H$\beta$
as a function of oxygen abundance 12 + log(O/H). The dots show individual 
data points 
 while the open circles show the means 
of the data points in the intervals 7.1 - 7.6, 7.6 - 8.1 and 
8.1 - 8.6 of 12 + logO/H. The error bars show the mean error of the data 
points in each interval.
\label{Fig5}}

\end{figure}

\clearpage

\begin{figure}
\figurenum{6}
\plotfiddle{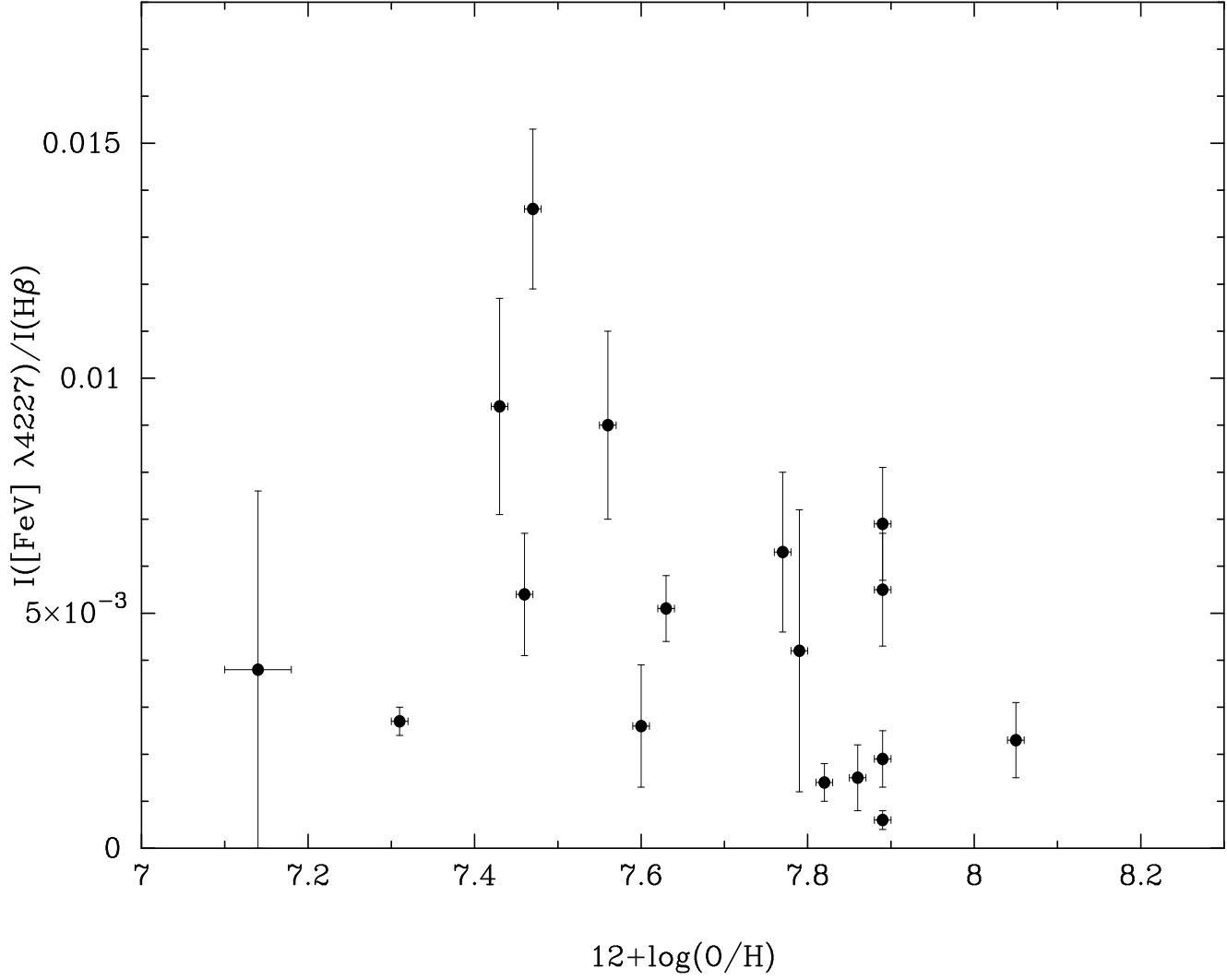}{1pt}{-90.}{400.}{500.}{-40.}{0.}
\caption{Intensity of the [Fe {\sc v}] $\lambda$4227
emission line relative to H$\beta$
as a function of oxygen abundance 12 + log(O/H).
\label{Fig6}}
\end{figure}

\clearpage

\begin{figure}
\figurenum{7}
\plotfiddle{Ne.ps}{1pt}{-90.}{400.}{500.}{-40.}{0.}
\caption{Intensity of the [Ne {\sc v}] $\lambda$3426
emission line relative to H$\beta$
as a function of oxygen abundance 12 + log(O/H).
\label{Fig7}}
\end{figure}

\clearpage

\begin{figure}
\figurenum{8}
\plotfiddle{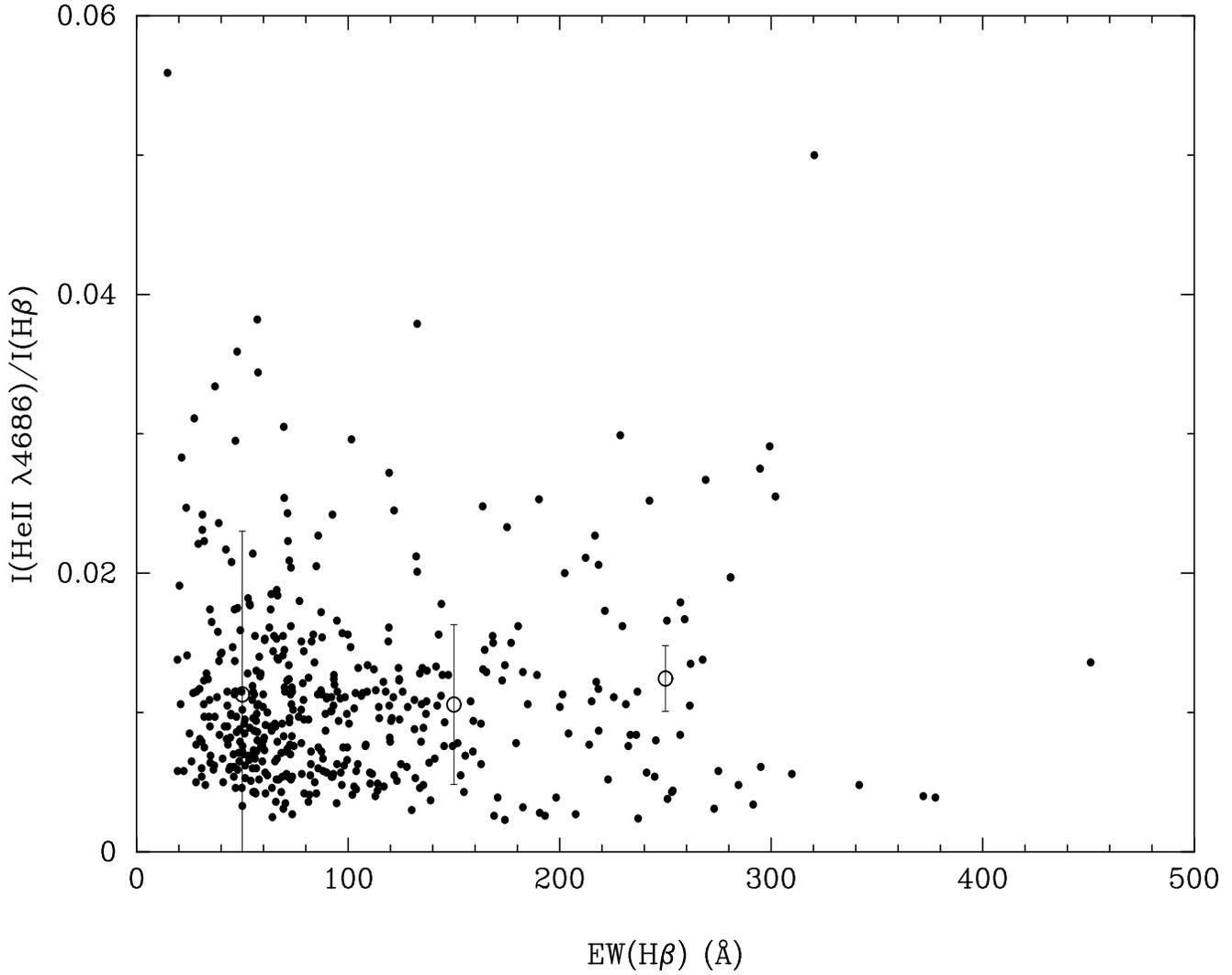}{1pt}{-90.}{400.}{500.}{-40.}{0.}
\caption{Intensity of the He {\sc ii} $\lambda$4686
emission line relative to H$\beta$
as a function of the equivalent width of H$\beta$.
The dots show individual data points 
 while the open circles show the means 
of the data points in the intervals 0 - 100 \AA, 100 - 200 \AA\ and 
200 - 300 \AA\ of EW(H$\beta$). The error bars show the mean error of the data 
points in each interval.
\label{Fig8}}
\end{figure}

\clearpage


\begin{figure}
\figurenum{9}
\epsscale{0.4}
\plotone{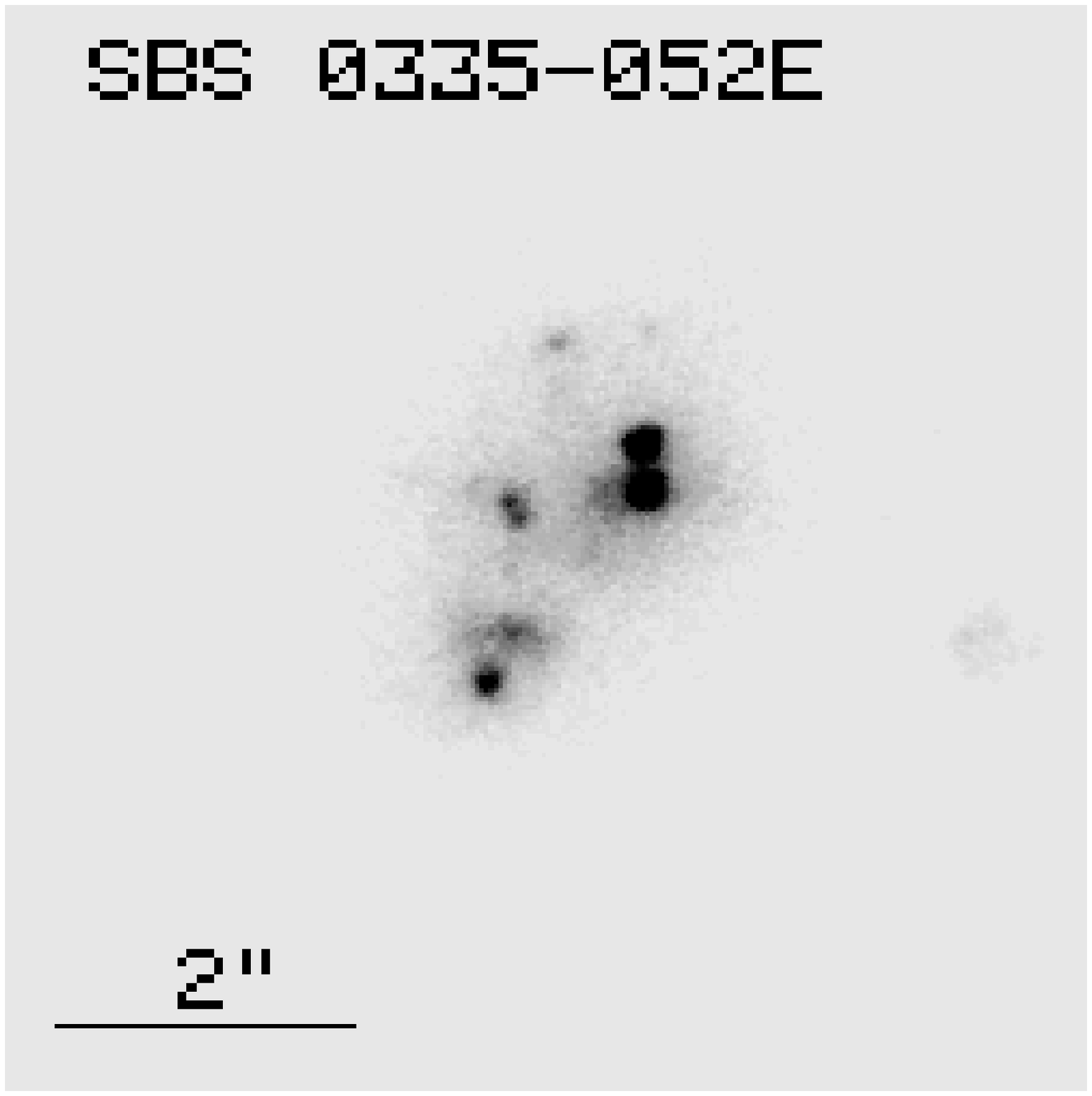}
\plotone{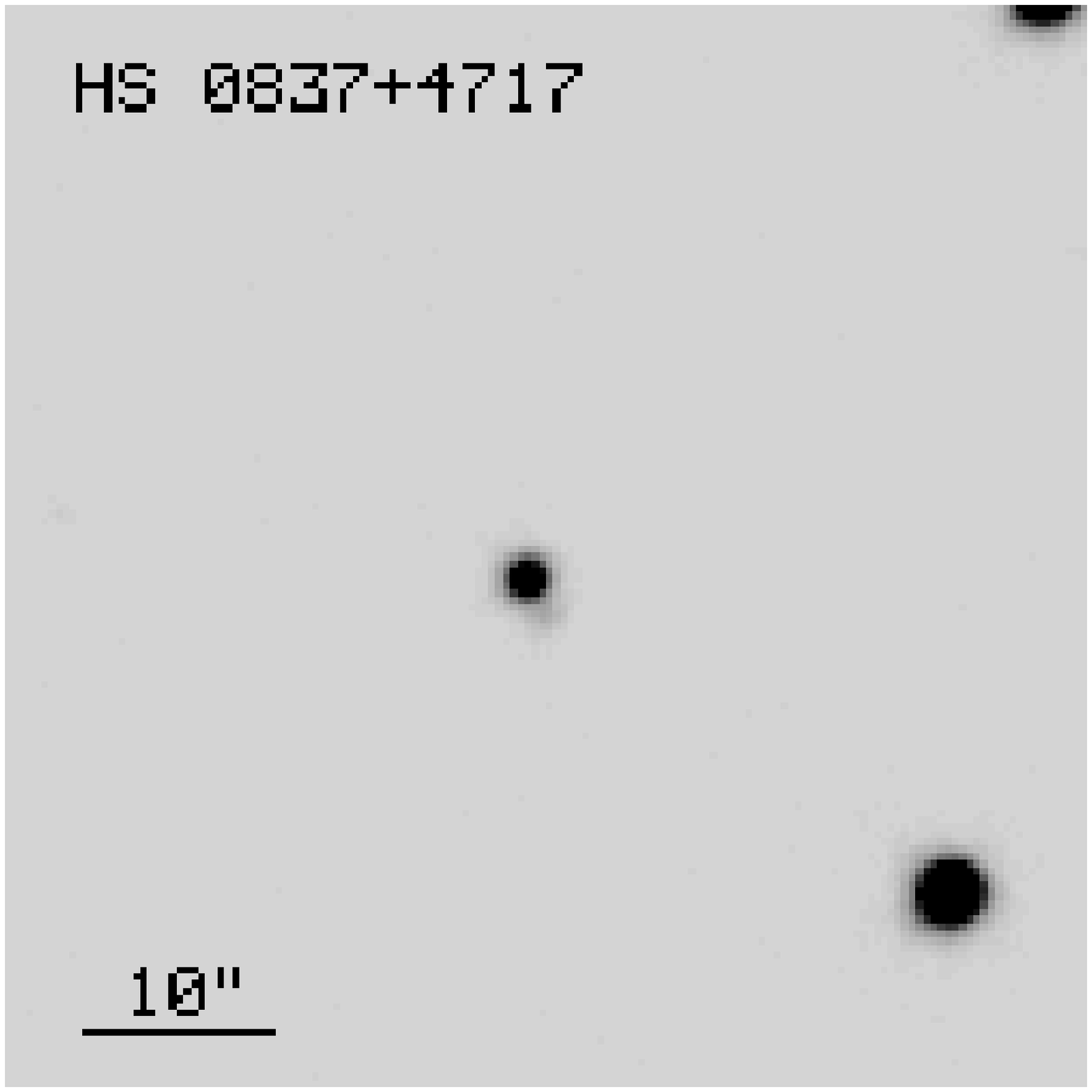}
\plotone{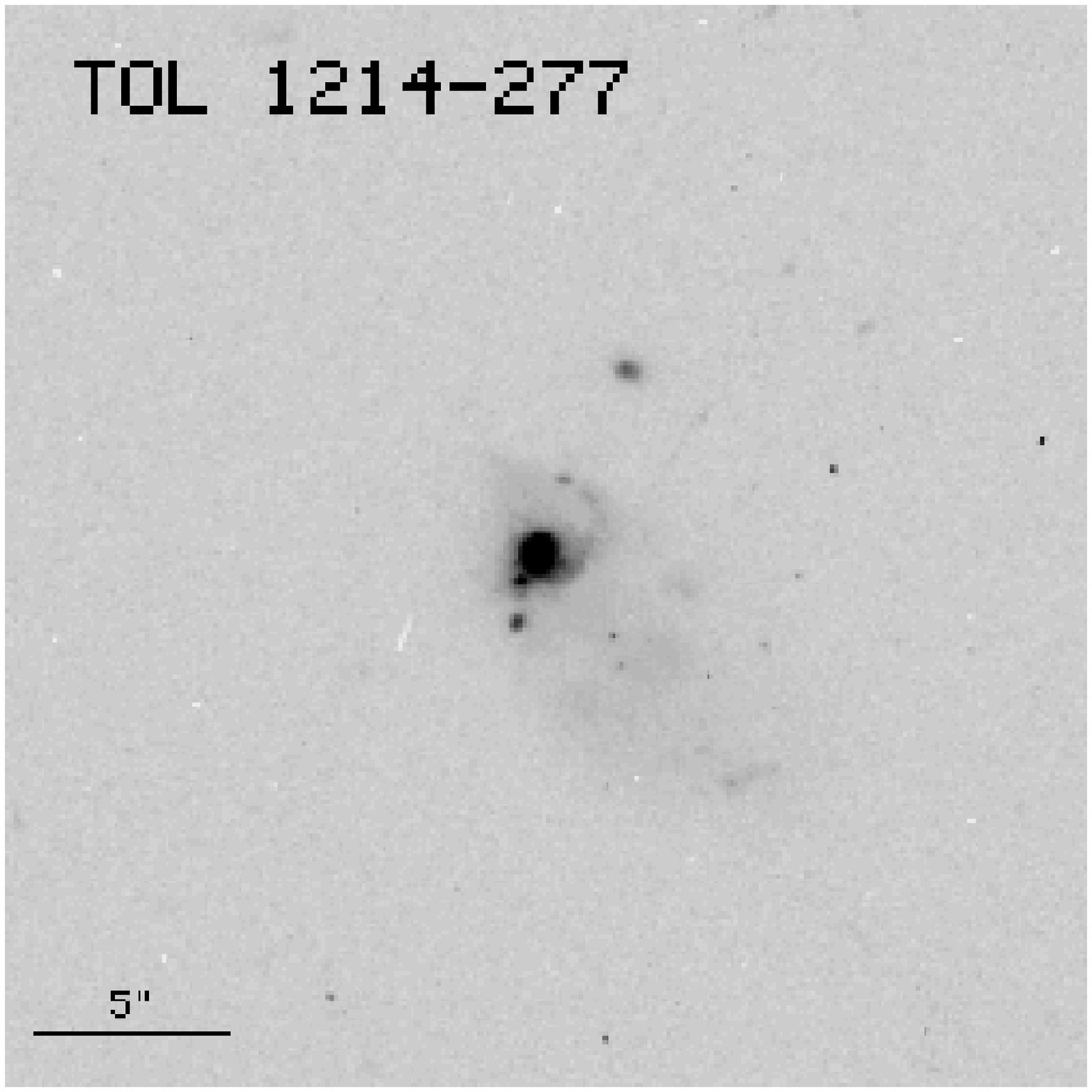}
\plotone{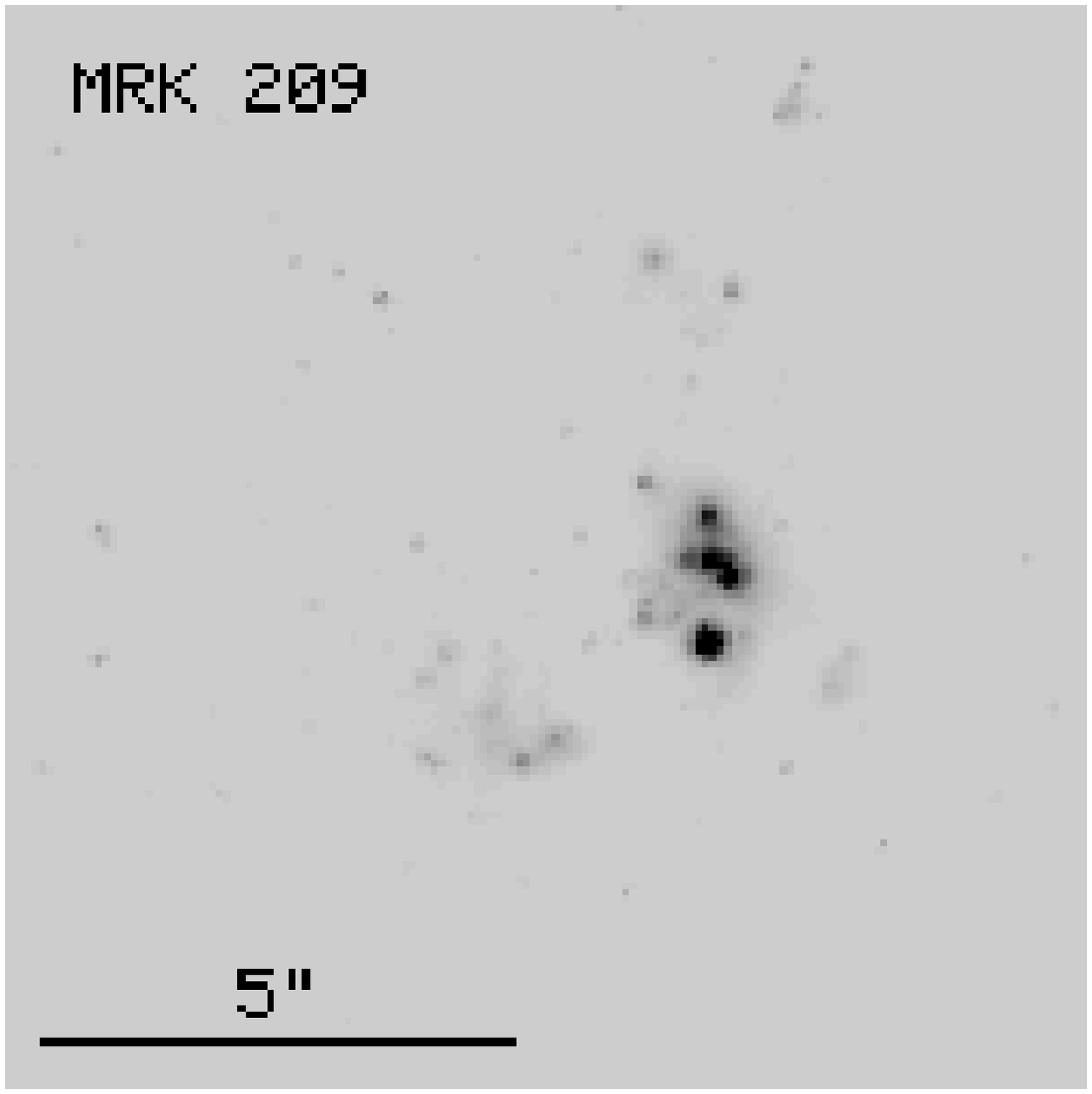}
\caption{Images of known galaxies with detected [Ne {\sc v}] emission
(see their spectra in Fig. \ref{Fig4}). North is at top and East to the left.
The angular scales are given by horizontal bars.
The pictures of SBS0335--052E, Tol 1214--277 and Mrk 209 are {\sl HST} images
taken respectively by the ACS camera with the F140LP filter, the WFPC2 camera
with the F555W filter and the NICMOS camera with the F110W filter. The
$g$ - picture
of HS 0837+4717 is a ground-based image from the Sloan Digital Sky Survey,
hence its considerably worse angular resolution.
\label{Fig9}}
\end{figure}



\begin{figure}
\figurenum{10}
\epsscale{0.9}
\plotone{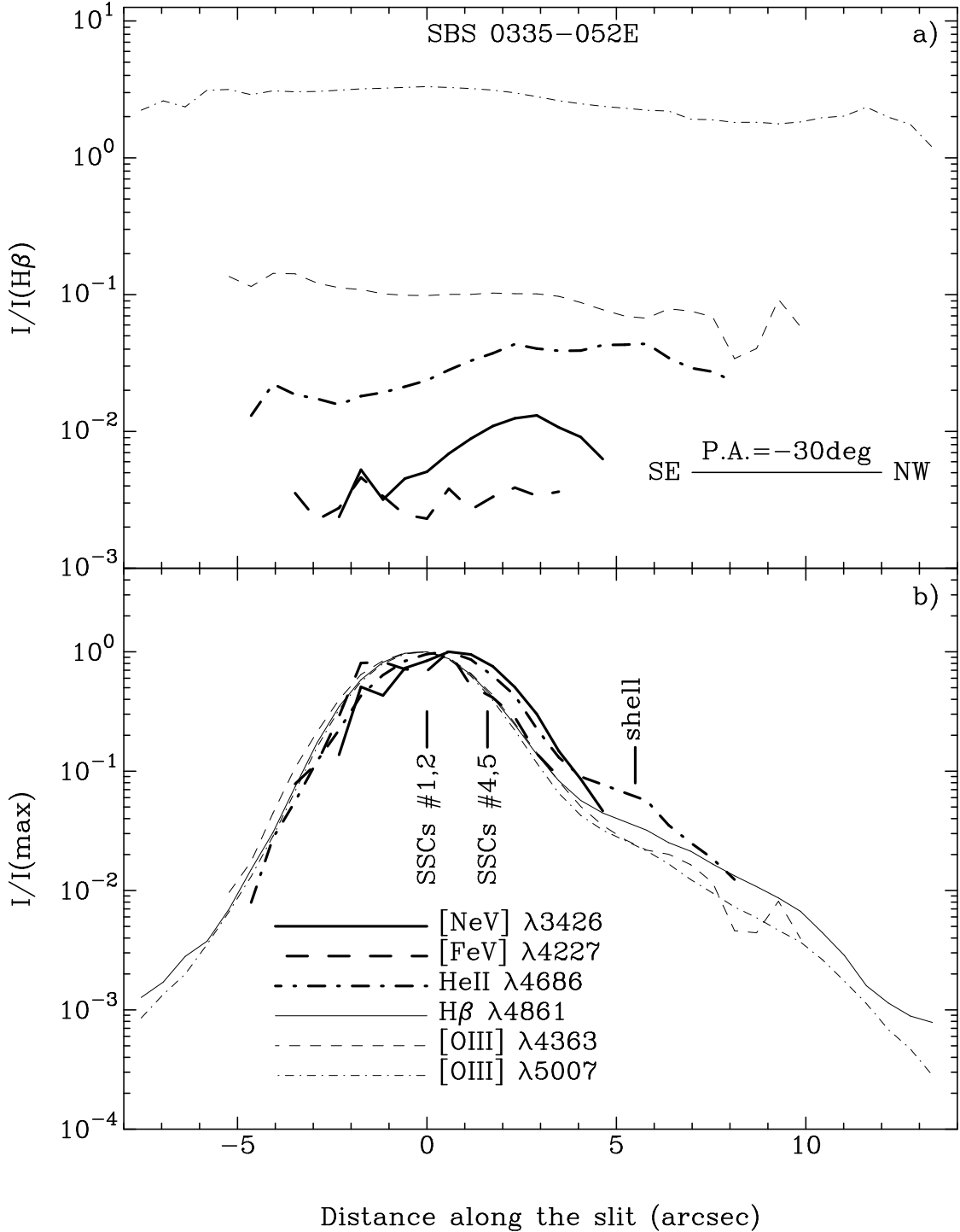}
\caption{Comparison of the spatial distributions along the position
angle of --30 degrees of the intensities of
various emission lines in SBS 0335--052E (a) relative to the intensity
of H$\beta$ and (b) relative to the maximum intensity of each line.
The location of SSCs \#1,2, SSCs \#4,5 and the shell \citep{T97} are indicated 
by short vertical lines. Note that the spatial distribution of the
[Ne {\sc v}] $\lambda$3426 emission line
is considerably more compact than those of other emission lines with softer
ionizing radiation, including He {\sc ii} $\lambda$4686. The latter shows 
an excess emission at the shell location.
\label{Fig10}}
\end{figure}

\begin{table}
\dummytable\label{Tab1}
\end{table}

\begin{table}
\dummytable\label{Tab2}
\end{table}

\begin{table}
\dummytable\label{Tab3}
\end{table}

\begin{table}
\dummytable\label{Tab4}
\end{table}

\begin{table}
\dummytable\label{Tab5}
\end{table}

\begin{table}
\dummytable\label{Tab6}
\end{table}

\begin{table}
\dummytable\label{Tab7}
\end{table}


\begin{thebibliography}{}

\bibitem[Abazajian et al. (2005)]{A05} Abazajian, K., et al. 2005, \aj,
129, 1755
\bibitem[Abel et al.(2002)]{A02} Abel, T., Bryan, G.~L., 
\& Norman, M.~L.\ 2002, Science, 295, 93 
\bibitem[Aller (1984)]{A84} Aller, L. H. 1984, Physics of Thermal Gaseous 
Nebulae  (Dordrecht: Reidel)
\bibitem[Bromm et al.(2002)]{B02} Bromm, V., 
Coppi, P.~S., \& Larson, R.~B.\ 2002, \apj, 564, 23
\bibitem[Campbell et al. (1986)]{C86} Campbell, A., Terlevich, R., \&  
Melnick, J. 1986, \mnras, 223, 811
\bibitem[Dopita \& Sutherland (1996)]{DS96} Dopita, M. A., \& 
Sutherland, R. S. 1996, \apjs, 102, 161
\bibitem[Feibelman (1996)]{F96} Feibelman, W. A., Hyung, S., \& 
Aller, L. H. 1996, \mnras, 278, 625
\bibitem[Ferland et al. (1998)]{F98} Ferland, G. J., Korista, K. T., 
Verner, D. A., Ferguson, J. W., Kingdon, J. B., \& Verner, E. M. 1998, 
\pasp, 110, 761
\bibitem[Fricke et al.(2001)]{F01} Fricke, K. J., Izotov, Y. I., 
Papaderos, P., Guseva, N. G., \& Thuan, T. X. 2001, \aj, 121, 169
\bibitem[Garnett (1992)]{G92} Garnett, D. R. 1992, \aj, 103, 1330
\bibitem[Garnett et al. (1991)]{G91} Garnett, D. R., Kennicutt, R. C., 
Chu, Y.-H., \& Skillman, E. D. 1991, \apj, 373, 458
\bibitem[Guseva et al. (2000)]{GIT00} Guseva, N. G., Izotov, Y. I., \& 
Thuan, T. X. 2000, \apj, 531, 776
\bibitem[Hirashita \& Hunt (2004)]{H04} Hirashita, H., \& Hunt, L.K. 2004, 
\aap, 421, 555
\bibitem[Izotov \& Thuan (1998)]{IT98} Izotov, Y. I., \& Thuan, T. X. 1998,
\apj, 497, 227 
\bibitem[Izotov \& Thuan (1999)]{IT99} Izotov, Y. I., \& Thuan, T. X. 1999, 
\apj, 511, 639
\bibitem[Izotov \& Thuan (2004a)]{IT04a} Izotov, Y. I., \& Thuan, T. X. 2004a, 
\apj, 602, 200
\bibitem[Izotov \& Thuan (2004b)]{IT04b} Izotov, Y. I., \& Thuan, T. X. 2004b, 
\apj, 616, 768
\bibitem[Izotov et al. (2001a)]{ICG01} Izotov, Y. I., Chaffee, F. H., \& 
Green, R. F. 2001a, \apj, 562, 727
\bibitem[Izotov et al. (2001b)]{ICS01} Izotov, Y. I., Chaffee, F. H., \& 
Schaerer, D. 2001b, \aap, 378, L45
\bibitem[Izotov et al. (2001c)]{I01c} Izotov, Y. I., Chaffee, F. H., 
Foltz, C. B., et al. 2001c, \apj, 560, 222
\bibitem[Izotov et al. (1994)]{ITL94} Izotov, Y. I., Thuan, T. X., \& 
Lipovetsky, V. A. 1994, \apj, 435, 647 
\bibitem[Izotov et al. (1997a)]{ITL97} Izotov, Y. I., Thuan, T. X., \& 
Lipovetsky, V. A. 1997a, \apjs, 108, 1
\bibitem[Izotov et al. (1997b)]{I97b} Izotov, Y. I., Foltz, C. B.,
Green, R. F., Guseva, N. G., \& Thuan, T. X. 1997b, \apj, 487, L37
\bibitem[Izotov et al. (1997c)]{I97c} Izotov, Y. I., Lipovetsky, V. A.,
Chaffee, F. H., Foltz, C. B., Guseva, N. G., \& Kniazev, A. Y. 
1997c, \apj, 476, 698
\bibitem[Izotov et al. (1999)]{I99} Izotov, Y. I., Chaffee, F. H., 
Foltz, C. B., Green, R. F., Guseva, N. G., \& Thuan, T. X. 1999, \apj, 527, 757
\bibitem[Izotov et al. (2004a)]{I04a} Izotov, Y. I., Noeske, K. G., 
Guseva, N. G., Papaderos, P., Thuan, T. X., \& Fricke, K. J. 2004a,
\aap, 415, L27
\bibitem[Izotov et al. (2004b)]{I04b} Izotov, Y. I., Stasi\'nska, G., 
Guseva, N. G., \& Thuan, T. X. 2004b, \aap, 415, 87
\bibitem[Izotov et al. (2004c)]{I04c} Izotov, Y. I., Papaderos, P., 
Guseva, N. G., Fricke, K. J., \& Thuan, T. X. 2004c, \aap, 421, 539
\bibitem[Izotov et al. (2005a)]{I05} Izotov, Y. I., Stasi\'nska, G., 
Meynet, G., Guseva, N. G., \& Thuan, T. X. 2005a, \aap, submitted
\bibitem[Izotov et al. (2005b)]{ITG05} Izotov, Y. I., Thuan, T. X., \&
Guseva, N. G. 2005b, \apj, in press; preprint astro-ph/0506498
\bibitem[Kennicutt et al. (1994)]{K94} Kennicutt, R.C., Tamblyn, O., 
\& Congdon, C.E. 1994, \apj, 435, 22
\bibitem[Kunth et al. (2003)]{K03} Kunth, D., Leitherer, C., Mas-Hesse, J. M.,
\"Ostlin, G., \& Petrosian, A. 2003, \apj, 597, 263
\bibitem[Kurucz (1979)]{K79} Kurucz, R. L. 1979, \apjs, 40, 1
\bibitem[Lipovetsky et al. (1999)]{L99} Lipovetsky, V. A., Chaffee, F. H., 
Izotov, Y. I., et al. 1999, \apj, 519, 177
\bibitem[Masegosa et al. (1994)]{M94} Masegosa, J., Moles, M., \& 
Campos-Aguilar, A. 1994, \apj, 420, 576                    
\bibitem[Oke (1990)]{O90} Oke, J. B. 1990, \aj, 99, 1621
\bibitem[Pustilnik et al. (2001)]{P01} Pustilnik, S., Brinks, E., 
Thuan, T. X., Lipovetsky, V. A., \& Izotov, Y. I. 2001, \aj, 121, 1413
\bibitem[Pustilnik et al. (2004)]{P04} Pustilnik, S., Kniazev, A., 
Pramskij, A., Izotov, Y., Foltz, C., Brosch, N., Martin, J.-M.,
\& Ugryumov, A. 2004, \aap, 419, 469
\bibitem[Schaerer (1996)]{S96a} Schaerer, D. 1996, \apj, 467, L17
\bibitem[Schaerer (2002)]{S02} Schaerer, D. 2002, \aap, 382, 28
\bibitem[Schaerer (2003)]{S03} Schaerer, D. 2003, \aap, 397, 527
\bibitem[Schaerer \& de Koter (1997)]{S97} Schaerer, D, \& de Koter, A. 
1997, \aap,322, 598
\bibitem[Schaerer \& Vacca (1998)]{SV98} Schaerer, D., \& Vacca, W. D. W. 
1998, \apj, 497, 618
\bibitem[Schulte-Ladbeck et al. (2001)]{S01} Schulte-Ladbeck, R. E., Hopp, U.,
Greggio, L., Crone, M. M., \& Drozdovsky, I. O. 2001, \aj, 121, 3007
\bibitem[Steidel et al. (1996)]{St96} Steidel, C. C., Giavalisco, M., 
Pettini, M., Dickinson, M., \& Adelberger, K. L. 1996, \apj, 462, 17
\bibitem[Stasi\'nska (1990)]{S90} Stasi\'nska G. 1990, \aaps, 83, 501
\bibitem[Stasi\'nska \& Izotov (2003)]{SI03} Stasi\'nska, G., \&
Izotov, Y. I. 2003, \aap, 397, 71
\bibitem[Terlevich et al. (1991)]{T91} Terlevich, R., Melnick, J., 
Masegosa, J., Moles, M., \& Copetti, M. V. F. 1991, \aaps, 91, 285
\bibitem[Thuan (1983)]{T83} Thuan, T.X. 1983, \apj, 268, 667
\bibitem[Thuan \& Izotov (1997)]{TI97} Thuan, T. X., Izotov, Y. I. 1997, 
\apj, 489, 623 
\bibitem[Thuan et al. (1995)]{TIL95} Thuan, T. X., 
Izotov, Y. I., \& Lipovetsky, V. A. 1995, \apj, 445, 108
\bibitem[Thuan et al. (1997)]{T97} Thuan, T. X., Izotov, Y. I., \&
Lipovetsky, V. A. 1997, \apj, 477, 661 
\bibitem[Thuan et al. (1999)]{T99} Thuan, T. X., Izotov, Y. I., \& 
Foltz, C. B. 1999, \apj, 525, 105
\bibitem[Thuan et al. (2004)]{T04} Thuan, T. X., Bauer, F. E., 
Papaderos, P., \& Izotov, Y. I. 2004, \apj, 606, 213
\bibitem[van Paradijs \& McClintock (1995)]{PM95} van Paradijs, J., \& McClintock, J.E. 1995, in X-ray Binaries,
ed. W.H. Lewin, J. van Paradijs, \& E.P.J. van den Heuvel 
(Cambridge: Cambridge Univ. Press), 58
\bibitem[van Zee et al. (2000)]{Z98} van Zee, L., Westpfahl, D., 
Haynes, M. P., \& Salzer, J. J. 1998, \aj, 115, 1000
\bibitem[Vilchez \& Pagel (1988)]{VP88} V\'{\i}lchez, J. M., \& 
Pagel, B. E. J. 1988, \mnras, 231, 257
\bibitem[Whitford (1958)]{W58} Whitford, A. E. 1958, \aj, 63, 201


\end{thebibliography}
\end{document}